\RequirePackage{fix-cm}
\documentclass[12pt]{article}
\usepackage{amsmath}
\usepackage[type1]{libertineRoman} \usepackage[scaled=0.92]{helvet}
\usepackage[libertine]{newtxmath}

\usepackage[T1]{fontenc}
\usepackage[latin9]{inputenc}
\usepackage{geometry}
\usepackage{color}
\usepackage{array}
\usepackage{verbatim}
\usepackage{float}
\usepackage{url}
\usepackage{setspace}
\usepackage[authoryear]{natbib}
\PassOptionsToPackage{normalem}{ulem}
\usepackage{ulem}
\usepackage[unicode=true]{hyperref}
\usepackage{breakurl}

\usepackage{multirow}
\usepackage{booktabs}

\makeatletter

\providecommand{\tabularnewline}{\\}

\usepackage{fancyhdr}

\providecommand{\tabularnewline}{\\}
\usepackage{abstract}
\usepackage{longtable}
\usepackage{rotating}

\newtheorem{theorem}{Theorem}

\newtheorem{definition}[theorem]{Definition}

\newtheorem{lemma}{Lemma}

\newtheorem{proposition}{Proposition}

\usepackage{lscape}

\definecolor{darkblue}{rgb}{0.00,0.15,0.7}

\hypersetup{
    colorlinks=true,		    linkcolor=darkblue,		    citecolor=black, 		    urlcolor=darkblue,
    bookmarksopen=true,
    pdfpagemode=UseNone,     pdftitle={Verifying the existence of maximum likelihood estimates for generalized linear models},
    pdfauthor={Sergio Correia and Paulo Guimaraes and Tom Zylkin},
    pdfstartview= Fit,     pdftoolbar= true,
    pdfcenterwindow=true,
}

\usepackage{multicol}

\newcommand*{\QED}{\hfill\ensuremath{\blacksquare}\medskip{}}
\newcommand{\Regressors}{\ensuremath{X=x_1, x_2, \ldots, x_M}\xspace}
\usepackage{xspace}

\usepackage{dsfont}
\usepackage{bbm} 

\makeatother

\begin{document}
\pagenumbering{gobble} 

\title{\textbf{\Large{}Verifying the existence of maximum likelihood estimates
 for generalized linear models}\thanks{We would like to thank
 Paul Kvam, Jo\~ao Santos Silva, Molin Zhong, and conference participants at the 2019 North American Summer Meeting of the Econometric Society and the 2019 Stata Conference. Tom Zylkin is grateful for research support from the NUS Strategic Research Grant (WBS: R-109-000-183-646)
 awarded to the Global Production Networks Centre (GPN@NUS) for the project titled
 ``Global Production Networks, Global Value Chains, and East Asian Development''. All data and examples described in the paper are available via an accompanying \href{https://github.com/sergiocorreia/ppmlhdfe/blob/master/guides/README.md}{website}.  }{\Large{} }}

\author{\noindent \hspace{0em}\begin{tabular}{c}
Sergio Correia{\vspace{-2mm}
}\tabularnewline
\emph{Federal Reserve Bank of Richmond}\tabularnewline
\end{tabular}\and\hspace{-3em}\begin{tabular}{c}
Paulo Guimar\~aes{\vspace{-2mm}
}\tabularnewline
\emph{Banco de Portugal and Universidade do Porto}\tabularnewline
\end{tabular} \vspace{0cm}
\and\hspace{-2em} \begin{tabular}{c}
Tom Zylkin\thanks{\emph{Corresponding author}: Tom Zylkin. Robins School of Business,
University of Richmond, Richmond, VA, USA 23217. E-mail: \protect\url{tzylkin@richmond.edu}.}{\footnotesize{}\vspace{-2mm}
}\tabularnewline
\emph{University of Richmond}\tabularnewline
\end{tabular}}

\date{\medskip{}\today}
\maketitle
\begin{abstract}
\begin{singlespace} A fundamental problem with nonlinear models is that maximum likelihood estimates are not guaranteed to exist. Though nonexistence is a well known problem in the binary response model literature, it presents significant challenges for other models and is not as well understood in more general settings. These challenges are only magnified for models that feature many fixed effects and other high-dimensional parameters. We address the current ambiguity surrounding this topic by studying the conditions that govern the existence of estimates for (pseudo-)maximum likelihood estimators used to estimate a wide class of generalized linear models (GLMs). We show that some, but not all, of these GLM estimators can still deliver consistent estimates of at least some of the linear parameters when these conditions fail to hold. We also demonstrate how to verify these conditions in models with high-dimensional parameters, such as panel data models with multiple levels of fixed effects.
Applying our methods to a gravity model with heterogeneous free trade agreement effects, we show that failing to detect nonexistence can produce misleading numerical estimates.
\end{singlespace}
\end{abstract}
\noindent \vfill{}

\noindent
\thispagestyle{empty}

\newpage{}
\pagenumbering{arabic} 

\section{Introduction}

\label{section_introduction}\setcounter{page}{1} Estimators based
on count data models are widely used in applied economic research
(\citealp{cameron_regression_2013}; \citealp{winkelmann_econometric_2013}).
In particular, Poisson regression has exploded in popularity since
the publication of \citet{santos_silva_log_2006}.\footnote{Poisson pseudo-maximum likelihood (PML) estimators have emerged as
a workhorse approach for studying health outcomes (\citealp{manning_estimating_2001}),
patent citations (\citealp{figueiredo_industry_2015}), trade (\citealp{orourke_aer2019}),
migration (\citealp{bertoli_rational_2019}), commuting (\citealp{brinkman2019freeway}),
auctions (\citealp{bajari_rand2003}), finance (\citealp{cohn2021count}),
and many other economic applications with nonnegative dependent variables
and/or with data assumed to be generated by a constant-elasticity
model.} Given this widespread and longstanding popularity, it is genuinely
surprising that economists have only become aware relatively recently
that count data models are not guaranteed to have maximum likelihood
(ML) solutions. More precisely, \citet{santos_silva_existence_2010} show that the
first-order conditions that maximize the likelihood of Poisson models
might not have a solution if regressors are perfectly collinear over
the subsample where the dependent variable is nonzero. Beyond this
observation, however, \citet{santos_silva_existence_2010} caution
that ``it is not possible to provide a sharp criterion determining
the existence'' of Poisson ML estimates. Moreover, although nonexistence
is a well known issue in binary outcome models, it seemingly remains
unknown if similar issues could arise in other nonbinary outcome models
besides Poisson, and the connections between these various cases remain seemingly
unknown as well.

In this paper, we resolve several key aspects of this ambiguity. {We
document that} nonexistence of ML estimates is a potential problem
for a broad class of generalized linear models (GLMs), including Poisson,
binary outcome models such as probit and logit, as well as several
other models. {So as not to overstate our own theoretical
contribution, the former result is taken from \citet{verbeek1989compactification},
and similar results for related classes of models can be found in \citet{aickin1979existence},
\citet{geyer1990likelihood}, \citet{clarkson_computing_1991},
\citet{geyer2009likelihood}, and \citet{fienberg2012maximum}. As we discuss below, these results are
often not given enough attention. }We also clarify that this problem
continues to be salient for pseudo-maximum likelihood (PML) estimators
of these models and, furthermore, that some common PML estimators
are affected by nonexistence in ways that cannot be remedied without
changing the estimator. For cases in which simpler remedies are possible,
we discuss computational methods for detecting and resolving nonexistence
and propose a novel algorithm that works well even in settings that
require a complex array of high-dimensional covariates, such as panel
data models with multiple levels of fixed effects.

We derive our main results in part by drawing on a largely uncredited
contribution by \citet{verbeek1989compactification}, who established
necessary and sufficient conditions governing the existence of ML
estimates for a broad class of GLMs.\footnote{As of this writing, both printings of \citet{verbeek1989compactification,verbeek_compactification_1992}
together have only eleven unique citations listed on Google Scholar. Another notable earlier contribution by \citet{aickin1979existence}, discussed below,
appears to be similarly uncredited.
} 
{} Using Verbeek's earlier results as our starting point, we show that
for many GLMs, even when the ML estimates can nominally be said to
``not exist'', at least some of the linear parameters can usually
be consistently estimated. We also add new results for PML estimation approaches that have only
become popular {in more recent years} and that turn
out not to share these useful properties. For example, the log-link
gamma PML estimator sometimes recommended in fields such as international
trade and health care economics has very different conditions governing
nonexistence than Poisson and suffers from more dire consequences
when it occurs. 

In addition to discussing how to detect such a problem, we also provide
guidance on what can be done about it. At the moment, this is another
area in need of clarity. Even for binary response models, where nonexistence
is well known as the so-called ``separation'' problem, textbooks
that mention the topic generally stop short of suggesting remedies
(\citealp{zorn_solution_2005,eck2018computationally}). The binary
outcome model literature has filled this gap primarily by presenting
a choice between two main ways of solving the problem, each with its
limitations. On the one hand, the most common approach is to drop
a regressor from the model (\citealp{zorn_solution_2005,allison2008convergence,rainey2016dealing}). This is also the approach that has been discussed most often in the context
of models for nonbinary outcomes (\citealp{santos_silva_existence_2010,larch_currency_2017}).
On the other hand, dropping a regressor has implications for the estimation
and identification of the other parameters, and often it is not obvious
which regressor is the ``right'' one to drop. Thus, a leading
alternative recommended for binary outcome settings is to impose a penalty on the likelihood function, often interpreted as assuming the
parameters have been drawn from a known prior distribution (\citealp{heinze2002solution,gelman2008weakly}).
These methods can be adapted to settings with nonbinary outcomes
as well (\citealp{firth1993bias,kosmidis2020mean}), and they have the advantage
that they can produce finite estimates for all of the model parameters even when separation occurs.
{However,
because they modify the objective being maximized, the estimates they yield
are not ML estimates and thus are not directly comparable to those from
the original unpenalized GLMs.}
Furthermore, they are
not currently compatible with models that include high-dimensional
fixed effects, which are widely used in the international trade literature
(\citealp{head_gravity_2014,yotov_advanced_2016}) and are becoming
increasingly popular in applied work in general.\footnote{{
The term high-dimensional fixed effects refers to the inclusion of multiple sets of fixed effects (e.g., individual, firm, time), where at least one set contains so many categories that estimating the model with dummy variables directly is computationally impractical.} The popularity of this type of model is only likely to increase in the near future thanks
to a series of computational innovations that have made models with
multiple levels of fixed effects more feasible to compute (see \citealp{figueiredo_industry_2015,larch_currency_2017,stammann2017fast,berge2018efficient,ppmlhdfe})
as well as a growing literature on bias corrections for incidental
parameter bias (see \citealp{arellano_understanding_2007,fernandez-val_individual_2016}).}

Our own {preferred} remedy,
which {in
practice} only involves withholding the separated observations from
the estimation sample, is generally very simple to implement {through
the new algorithms we introduce} and does not have any of these limitations.\footnote{Our recommendation to withhold separated observations from the estimation is ostensibly similar
to \citet{allison2008convergence}'s suggestion to ``do nothing'',
as doing nothing could result in approximately valid estimates and
inferences for at least some of the model parameters. However, in
general, doing nothing could result in lack of
numerical convergence or---in the worst case---convergence to incorrect
values. In the words
of \citet{geyer2019slides}, ``no one knows how much applied statistics
is garbage because of this.'' Also, some software packages drop separated
observations by default (e.g., Stata's \texttt{probit} command), but
they generally are not adept at detecting these observations, nor
do they usually provide theoretical justification for this practice
in their documentation. Our companion \href{https://github.com/sergiocorreia/ppmlhdfe/blob/master/guides/README.md}{website} offers examples and discussion; see \url{github.com/sergiocorreia/ppmlhdfe/blob/master/guides}.}
{The theoretical justification for withholding these
observations} comes from an insight advanced independently by \citet{aickin1979existence},
\citet{verbeek1989compactification},
\citet{geyer1990likelihood}, and \citet{clarkson_computing_1991}:
a model suffering from separation can often be nested within a ``compactified''
model where the conditional mean of each observation is allowed to go
to its boundary values. The (pseudo-)likelihood function always has
a maximum somewhere in the compactified parameter space; thus, we
can transform the problem of nonexistence to one of possible corner
solutions. More importantly, observations with a conditional mean
at the boundary in the more compactified model are effectively perfectly
predicted observations. These observations offer no information about
the parameters with interior solutions and, as we will show, can be
quickly detected even for very complex models. Removing these observations
then results in a standard (non-compactified) version of the model
that is assured to produce the same model fit as the compactified
version, as well as the same point estimates and inferences of the
parameters with interior solutions. {As such, our approach
is equivalent in practice to fitting what \citet{geyer2009likelihood}
calls the ``limiting conditional model'' that conditions on the
separated observations. }We also show that {the estimates
for the estimable parameters} are consistent and that correct inference
requires only careful attention to which of the regressors are involved
in separation. The resulting output, on the whole, is no different
than what one would observe with a perfectly collinear regressor,
and the problems of interpretation and inference turn out to be very
similar as well.

{Although separation becomes equivalent to perfect collinearity after excluding the separated observations, the two concepts differ in important aspects. In the case of separation, all regressors are important to the fit of the model, including those without finite estimates. Furthermore, while the separated observations are withheld from the estimation step, the model yields predicted values for them that are consistent with fitting the model over the full sample. For models estimated via ML, it is also possible to obtain one-sided confidence bounds for the parameters whose estimates diverge to infinity. For discrete-response GLMs with canonical links, one can even obtain meaningful inferences on the predicted values of the separated observations (see \citealp{eck2018computationally}).}

{
Since the possible non-existence of a finite MLE is a well known problem in the context of binary outcome models, it may be surprising that equally fundamental results for Poisson regression and other nonbinary GLMs remain undercited by comparison.
This imbalance reflects both history and practice: the problem was easier to visualize geometrically for binary logit and probit models and more frequently encountered in applied research, leading to its widespread acknowledgment.
By contrast, the analogous results for Poisson and other nonbinary GLMs have mostly been circulated in the statistical theory literature.
This literature begins with \citet{haberman_log-linear_1973,haberman_analysis_1974}'s derivation of a necessary and sufficient condition for the existence of estimates for log-linear frequency table models. Notably, it was known at the time that this condition was difficult to verify for higher-dimensional tables (see \citealp{albert_existence_1984}), a still-unsettled problem we indirectly solve in this paper. Soon thereafter, \citet{wedderburn_existence_1976} independently derived a sufficient but not necessary condition for the existence of estimates across a wide class of GLMs that included Poisson and Gamma.\footnote{His result can be shown to be equivalent to \citet{santos_silva_existence_2010}'s later result for the Poisson model.} The first statement of a necessary and sufficient condition for the existence of Poisson regression estimates was given by \citet{aickin1979existence}, who derived results for models in the linear discrete exponential family.} 

{
\citet{silvapulle_existence_1981} and \citet{albert_existence_1984} are then credited with popularizing the concepts of ``separation'' and ``overlap'' for binary outcome models. \citet{silvapulle_existence_1981} gave these terms algebraic meaning, while \citet{albert_existence_1984} provided an influential geometric interpretation that distinguished between ``complete'' and ``quasi-complete'' separation. A few years later, \citet{silvapulle_existence_1986} showed how the conditions for separation studied in \citet{silvapulle_existence_1981} can be formulated as a linear programming problem, thus marking an important step towards detecting the issue in practice. Finally, \citet{verbeek1989compactification}, \citet{geyer1990likelihood}, and \citet{clarkson_computing_1991} independently extended these ideas to broader classes of models. Verbeek (1989) was the first to unify the conceptualizations of the problem that were being used in the binary outcome literature and the more general GLM setting, while \citet{geyer1990likelihood} and \citet{clarkson_computing_1991} developed similar insights for the linear exponential family and for ``models with a linear part'' (\citealp{stirling1984iteratively}), respectively.\footnote{{The ``models with a linear part'' concept studied in \citet{clarkson_computing_1991} is a broader category than GLMs, as it includes any other models where the parameters enter the likelihood function via a linear index function. The Tobit model for censored data is an example of a non-GLM model conforming to this framework.}}
Despite these advances, awareness of the possible nonexistence of estimates for nonbinary GLMs did not spread widely in applied work until it was highlighted in \citet{santos_silva_existence_2010}.
}

We add to this earlier literature in three main ways. First, by considering
an expanded set of estimation approaches, we offer a more detailed
treatment of how the separation problem varies across GLMs used {in
economics research} and estimators thereof. For example, a significantly
stricter set of conditions governs the existence of estimates for
gamma PML and inverse Gaussian PML than for Poisson, logit, and probit---a
result that raises concerns about applications of the former estimators
to settings where zero outcomes are common, such as health care cost
analysis and international trade. Importantly, these are precisely
the settings where these estimators have come into common usage; see,
e.g., \citet{manning_estimating_2001,egger_glm_2015}.\footnote{\citet{manning_estimating_2001} leave aside the issue of zero outcomes
in their paper, but indicate that gamma PML is generally a good
model for health care cost data and also remark that ``there is ostensibly
nothing in the above analysis that would preclude applications to
data where realizations of $y$ are either positive or zero, as is
common in many health economics applications.'' Our own findings
indicate that zeroes do pose a distinct problem for gamma PML estimation
that must be carefully taken into account.} Second, we clarify that at least some of the linear parameters can
be consistently estimated in the presence of separation as well as
how to obtain valid asymptotic inferences---though, again, it is
important to note these results do not extend to all of the estimators
we consider.\footnote{\citeauthor{gourieroux_pseudo_1984} (1984, Appx 1.1) and \citet{fahrmeir1985consistency}
(Sec. 2.2) both assume in their proofs of consistency that the solutions
for the linear parameters are interior. We present a proof that relies
on a suitable reparameterization of the separated model such that
the results of \citet{gourieroux_pseudo_1984} apply directly.}

Finally, we introduce a simple-but-powerful method for detecting separation
in models with a large number of fixed effects, a conceptually nontrivial
task that would ordinarily require solving a high-dimensional linear
programming problem. Because our algorithm relies on repeated iteration
of a least-squares-with-equality-constraints regression, it can take
advantage of the recent innovations of \citet{correia_linear_2017},
who shows how to solve high-dimensional least-squares problems in
nearly linear time. To our knowledge, the only other method that has
been suggested for detecting separation in large ML settings is that
of \citet{eck2018computationally}. {While their primary
focus is on obtaining inference {for the separated observations} when the ML estimate lies in the Barndorff-Nielsen
completion (\citealp{barndorff1978information}}), their method also detects separation by computing
the null eigenvectors of the Fisher information matrix, which reveals
the estimable and non-estimable components. In contrast, our methods
avoid large matrix operations altogether, which should make them substantially
more scalable. {In addition, their approach detects separation ex post after first
running the estimation algorithm on the full data set, while ours aims to detect separation
ex ante. {As such, our contribution concerns the task of detecting the separated observations rather than the downstream problem of how to do inference on them like in \citet{eck2018computationally}.} Due to its scalability, our approach can provide a solution for models
that fall outside the scope of currently available implementations for existing remedies,
including not only detection methods such as \citet{eck2018computationally} and \citet{Kosmidis2021detectseparation} but also
penalized likelihood methods such as \citet{gelman2008weakly} and \citet{kosmidis2020mean}.}

The rest of the paper proceeds as follows. Section 2 formally establishes
the problem of separation in GLMs, including its sufficient and necessary
conditions. Section 3 discusses how to address separation in setups
with and without fixed effects. Section 4 provides an empirical example.
Section 5 concludes. Further details are available in the Appendix,
including additional proofs and results of interest. We have also
created a \href{https://github.com/sergiocorreia/ppmlhdfe/blob/master/guides/README.md}{website} dedicated
to the separation problem, which provides numerous examples illustrating
the methods and principles described in this paper. {These examples include demonstrations for logit and multinomial
logit taken from the literature as well as 17 examples for Poisson and Poisson PML that we ourselves have curated.}

\section{Nonexistence in generalized linear models}  \label{section_separation}
The class of GLM-based estimators we consider is
defined by the maximization of the following log (pseudo-)likelihood objective function, corresponding to distributions of the exponential family:
\begin{align}
l\left(\beta\right) & =\sum_{i}l_{i}\left(\beta\right)=\sum_{i}\left[\alpha_{i}(\varphi)\,y_{i}\,\theta_{i}-\alpha_{i}(\varphi)\,b(\theta_{i})+c(y_{i},\varphi)\right],\label{eq:glm}
\end{align}
For brevity, we will generally refer to this objective function as the ``likelihood'', though we will use ``pseudo-likelihood'' when strictly discussing PML estimators. The individual term $\ell_i$ will be the ``likelihood contribution'' or ``pseudo-likelihood contribution''. $y_{i}\ge0$ is an outcome variable, $x_{i}$ is a set of $M$ regressors ($x_{1},x_{2},\ldots,x_{M}$), and $\beta\in\mathbb{R}^{M}$
is an $M\times1$ vector of parameters to be estimated. The function
$\alpha_{i}(\varphi)>0$ is usually of the form $w_{i}/\varphi$,
where $w_{i}$ is a known weight, and $\varphi$ is a potentially
unknown scale or dispersion parameter.\footnote{$\varphi$ is not associated with the problem of separation and will
henceforth be treated as known. The results we document apply to models
with unknown scaling factors without loss of generality. Table \ref{tab:Key-results-for}
gives examples. For more information on this class of models, see
\citet{McCullaghNelderGLM} Section 2.2.2.} $\theta_{i}=\theta(x_{i}\beta;\nu)$ is the canonical location parameter,
which links the linear predictor of a given observation $x_{i}\beta$ to its likelihood contribution $l_{i}$ and its conditional mean $\mu_{i}\equiv E[y_{i}|x_{i}]=b^{\prime}(\theta_{i})$.
Note that $\theta(x_{i}\beta;\nu)$ is continuous, strictly increasing,
and twice differentiable in $x_{i}\beta$ and that $b(\theta_{i})$
is continuous, increasing, and convex in $\theta_{i}$. Notably, these
last few restrictions together ensure that the quantities $\theta_{i}$,
$x_{i}\beta$, and $\mu_{i}$ are each increasing with respect to
one another and that $l\left(\beta\right)$ is continuous in $\beta$.
We further assume that $\lim_{x_{i}\beta\rightarrow-\infty}\mu_{i}=0$
to rule out the simple linear model, which always has a solution.\footnote{Also note that the linear predictor term $x_{i}\beta$ is often denoted
as $\eta_{i}$. We keep it as $x_{i}\beta$ to economize on notation.} $\nu$ is an additional dispersion parameter that allows us to also
consider negative binomial models (see Table \ref{tab:Key-results-for}).
Lastly, $c(\cdot)$ is a known real-valued function that depends on
the specific GLM.

\begin{sidewaystable}[p]
\caption{Mapping different regression models onto GLM \label{tab:Key-results-for}}
\medskip{}
\scalebox{.9}{\begin{tabular}{>{\raggedright}p{1.75cm}|>{\raggedright}p{7cm}|>{\raggedright}p{2.5cm}|>{\raggedright}p{3cm}|>{\raggedright}p{2.15cm}|>{\raggedright}p{6cm}}
Model & (Pseudo) Log-likelihood ($l$) & $\theta(x_{i}\beta;\nu)$ & $b(\theta_{i})$ & $\mu_{i}(=b^{\prime})$ & First-order condition for $\beta_{m}$\tabularnewline
\hline
Probit & $\sum_{i}\left(y_{i}\log\mu_{i}+\left(1-y_{i}\right)\log\left(1-\mu_{i}\right)\right)=$

$\sum_{i}\left(y_{i}\log\frac{\Phi_{i}\left(x_{i}\beta\right)}{1-\Phi_{i}\left(x_{i}\beta\right)}+\log\left(1-\Phi_{i}\left(x_{i}\beta\right)\right)\right)$ & $\log\frac{\Phi_{i}\left(x_{i}\beta\right)}{1-\Phi_{i}\left(x_{i}\beta\right)}$ & $\log\left(1+\exp\left(\theta_{i}\right)\right)$ & $\frac{\exp\left(\theta_{i}\right)}{1+\exp\left(\theta_{i}\right)}$

$(=\Phi(x_{i}\beta))$ & $\sum_{i}\frac{\phi\left(x_{i}\beta\right)}{\Phi_{i}\left(x_{i}\beta\right)\left[1-\Phi_{i}\left(x_{i}\beta\right)\right]}\left[y_{i}-\mu_{i}\right]x_{mi}=0$\tabularnewline
\hline
Logit & $\sum_{i}\left(y_{i}\log\mu_{i}+\left(1-y_{i}\right)\log\left(1-\mu_{i}\right)\right)=$

$\sum_{i}\left(y_{i}x_{i}\beta-\log\left(1+\exp\left(x_{i}\beta\right)\right)\right)$ & $x_{i}\beta$ & $\log\left(1+\exp\left(\theta_{i}\right)\right)$ & $\frac{\exp\left(\theta_{i}\right)}{1+\exp\left(\theta_{i}\right)}$ & $\sum_{i}\left[y_{i}-\mu_{i}\right]x_{mi}=0$\tabularnewline
\hline
Poisson & $\sum_{i}\left[y_{i}x_{i}\beta-\exp\left(x_{i}\beta\right)-\ln y_{i}!\right]$ & $x_{i}\beta$ & $\exp\left(\theta_{i}\right)$ & $\exp\left(\theta_{i}\right)$ & $\sum_{i}\left[y_{i}-\mu_{i}\right]x_{mi}=0$\tabularnewline
\hline
Negative Binomial & $\sum_{i}y_{i}\log\left(\frac{\exp\left(x_{i}\beta\right)}{\nu+\exp\left(x_{i}\beta\right)}\right)-\nu\log\left(\nu+\exp\left(x_{i}\beta\right)\right)+c\left(\nu,y_{i}\right)$ & $\log\left(\frac{\exp\left(x_{i}\beta\right)}{\nu+\exp\left(x_{i}\beta\right)}\right)$ & $\nu\log\left(\frac{\nu}{1-\exp(\theta)}\right)$ & $\nu\frac{\exp\left(\theta_{i}\right)}{1-\exp\left(\theta_{i}\right)}$

$(=e^{x_{i}\beta})$ & $\sum_{i}\left[y_{i}-\mu_{i}\right]\left(1+\nu^{-1}\mu_{i}\right)^{-1}x_{mi}=0$\tabularnewline
\hline
Gamma (PML) & $\sum_{i}-\alpha y_{i}\exp\left(-x_{i}\beta\right)-\alpha x_{i}\beta$ & $-\exp\left(-x_{i}\beta\right)$ & $\log\left(-1/\theta_{i}\right)$ & -$1/\theta_{i}$

$(=e^{x_{i}\beta})$ & $\alpha\sum_{i}\left[y_{i}-\mu_{i}\right]\exp\left(-x_{i}\beta\right)x_{mi}=0$\tabularnewline
\hline
Gaussian & $\sum_{i}-\frac{1}{2\sigma^{2}}\left[y_{i}-\exp\left(x_{i}\beta\right)\right]^{2}-\frac{1}{2}\log\left(2\pi\sigma^{2}\right)=$

$\sum_{i}\frac{1}{\sigma^{2}}\left[y_{i}\exp\left(x_{i}\beta\right)-\frac{1}{2}\exp\left(2x_{i}\beta\right)\right]+c\left(\sigma^{2},y_{i}\right)$ & $\exp\left(x_{i}\beta\right)$ & $\theta_{i}^{2}/2$ & $\theta_{i}$ & $\frac{1}{\sigma^{2}}\sum_{i}\left[y_{i}-\mu_{i}\right]\exp\left(x_{i}\beta\right)x_{mi}=0$\tabularnewline
\hline
Inverse Gaussian (PML) & $\sum_{i}\alpha\left[-\frac{y_{i}}{2}\exp\left(-2x_{i}\beta\right)+\exp\left(-x_{i}\beta\right)\right]$ & $-\frac{\exp\left(-2x_{i}\beta\right)}{2}$ & $-(-2\theta)^{1/2}$ & $\left(-2\theta_{i}\right)^{-1/2}$

$(=e^{x_{i}\beta})$ & $\alpha\sum_{i}\left[y_{i}-\mu_{i}\right]\exp\left(-2x_{i}\beta\right)x_{mi}=0$\tabularnewline
\hline
\hline
\multicolumn{6}{>{\raggedright}p{23cm}}{$\Phi(\cdot)$ is the cdf of a standard normal distribution. $\phi(\cdot)$
is its pdf. $\alpha$ and $\sigma^{2}$ are dispersion/scaling factors
to be estimated, which do not affect identification of $\beta$. $\nu$,
which does affect identification of $\beta$, is the dispersion parameter
for the negative binomial regression. Note that for gamma and inverse
Gaussian, we consider only the pseudo-maximum Likelihood (PML) versions
of these estimators (the standard likelihood functions for these models
do not admit $y_{i}=0$ values.) These PML estimators each use a ``log link'' as opposed to the canonical link. We do the same for the Gaussian
GLM shown, since Gaussian (log link) PML is another common PML estimator.
The logit and probit likelihood functions can also be applied to fractional
data using Bernoulli PML; see \citet{papke_econometric_1996}.}\tabularnewline
\end{tabular}}
\end{sidewaystable}

The first-order condition for the $m$-th individual parameter, $\beta_{m}$, follows from the GLM score function:
\begin{equation}
s\left(\beta_{m}\right)=\sum_{i}s_{i}\left(\beta_{m}\right)=\sum_{i}\alpha_{i}(\varphi)\left[y_{i}-b^{\prime}(\theta_{i})\right]\theta^{\prime}(x_{i}\beta;\cdot)\,x_{mi}=0\quad\forall m.\label{eq:score}
\end{equation}
Examples of models conforming to this framework notably include binary
outcome models, count models such as Poisson and negative binomial,
and a variety of other closely related, non-GLM models such
as conditional logit, multinomial logit, and the Cox proportional-hazards model. In addition, as the score vectors
of many of these models can also be used to construct PML estimators
for continuous data, this framework also applies to
PML estimators such as  Poisson, gamma, Gaussian, inverse Gaussian,
and Bernoulli PML without loss of generality.
Note that our interest in PML estimators represents
an important deviation from \citet{verbeek1989compactification}
because PML estimation does not impose any restrictions on $c(y_{i},\varphi)$.
As such, we can consider potential nonexistence problems
in models where $y_{i}=0$ values would otherwise be inadmissible,
such as log-link gamma PML and other PML
estimators with similar score functions.\footnote{For more on the wide applicability of PML, see \citet{gourieroux_pseudo_1984}, \citet{manning_estimating_2001}, and \citet{santos_silva_log_2006}.}

On top of these general restrictions, we use two further assumptions
to derive a necessary and sufficient condition for existence that
holds across most of these estimators. First, we
assume that the matrix of regressors \Regressors is of full column
rank. This rank assumption allows us to set aside the more widely
understood case of perfectly collinear regressors, although in Section
\ref{section_solution}, we will find it useful to draw a comparison
between nonexistence and perfect collinearity.
Second, we assume for the moment that the individual likelihood
contributions $l_{i}(\beta)$ have a finite upper bound. Later on,
we will consider two estimators for which $\ell_i$ is not guaranteed to have a finite upper bound, gamma PML and inverse
Gaussian PML. We show that the relevant criteria governing existence
are not the same as when this assumption is met.

To extend and generalize the earlier result from \citet{santos_silva_existence_2010}
for Poisson models, we are now ready to prove the following proposition:

\begin{proposition} (Nonexistence) Suppose that $l(\beta)$ conforms
to \eqref{eq:glm}, the matrix of regressors \Regressors is of full
column rank, and the individual likelihood contribution $l_{i}(\beta)$ has a finite upper bound.
A solution for $\beta$ that maximizes \eqref{eq:glm} will \uline{not} exist if and only if there exists
a linear combination of regressors $z_{i}=x_{i}\gamma^{*}$ such that
\begin{align}
z_{i}=0\quad & \forall\,i\quad\text{s.t.}\quad0<y_{i}<\overline{y},\label{eq:inner}\\
z_{i}\ge0\quad & \forall\,i\quad\text{s.t.}\quad y_{i}=\overline{y},\label{eq:right}\\
z_{i}\le0\quad & \forall\,i\quad\text{s.t.}\quad y_{i}=0,\label{eq:left}
\end{align}
where $\gamma^{*}=(\gamma_{1}^{*},\gamma_{2}^{*},\ldots,\gamma_{M}^{*})\in\mathbb{R}^{M}$
is a nonzero vector of the same dimension as $\beta$ and where $\overline{y}$
is an upper bound on $\mu_{i}$ that equals $1$ for binary outcome
models ($\infty$ otherwise).\label{Prop1}\end{proposition}

The proof of this proposition follows \citet{verbeek1989compactification},
while also drawing on an earlier proof by \citet{silvapulle_existence_1981}
specifically for binary outcome models.\footnote{In \citet{verbeek1989compactification}, the relevant theorems are
Theorem 6, which establishes conditions under which the likelihood
function has a local maximum that lies on the boundary of the parameter
space, and Theorem 4, which establishes that any local maximum on
the boundary is a global maximum if the likelihood function is concave.
Note that we have relaxed the concavity assumption since it is straightforward
to show the weaker result that if there is a local
maximum at the boundary, the global maximum can only occur at the
boundary.} In addition, the necessity of the condition on the boundedness of
$l_{i}(\cdot)$ function is due to \citet{clarkson_computing_1991}; note that Proposition \ref{Prop2} later in the paper explores the implications of relaxing this assumption.

The general idea is that we want to show that if a vector $\gamma^{*}$
satisfying \eqref{eq:inner}-\eqref{eq:left} exists, then the likelihood
function $l(\beta)$ will always be increasing if we search for a maximum
in the direction associated with $\gamma^{*}$.\footnote{This is the same concept as what \citet{geyer2009likelihood} calls the ``generic direction of recession''. } Otherwise, if no such
$\gamma^{*}$ exists, then searching in any direction
from any starting point in $\mathbb{R}^{M}$ under the noted conditions will cause $l(\beta)$ to eventually decrease,
such that the function must reach a maximum for some finite $\beta^{MLE}\in\mathbb{R}^{M}$.

To proceed, let $\gamma=(\gamma_{1},\gamma_{2},\ldots,\gamma_{M})\in\mathbb{R}^{M}$
be an arbitrary nonzero vector of the same dimension as $\beta$
and let $k>0$ be a positive scalar. Now consider the function $l(\beta+k\gamma)$,
which allows us to consider how the likelihood
changes as we search in the same direction as $\gamma$ from
some initial point $\beta$. Differentiating $l(\beta+k\gamma)$ with
respect to $k$, we obtain
\begin{align*}
\frac{dl(\beta+k\gamma)}{dk} & =\sum_{i}\alpha_{i}(\varphi)\left[y_{i}-b^{\prime}(\theta_{i})\right]\theta_{i}^{\prime}x_{i}\gamma.
\end{align*}
Suppose there is a $\gamma^{*}$ such that $z_{i}=x_{i}\gamma^{*}$
 satisfies \eqref{eq:inner}-\eqref{eq:left}. In this case, setting $\gamma=\gamma^{*}$ the above
expression becomes
\begin{align*}
\frac{dl(\beta+k\gamma^{*})}{dk}=\sum_{y_{i}=0}\alpha_{i}(\varphi)\left[-b^{\prime}(\theta_{i})\right]\theta'z_{i}+\sum_{y_{i}=\overline{y}}\alpha_{i}(\varphi)\left[\overline{y}-b'(\theta_{i})\right]\theta'z_{i}>0,
\end{align*}
with the inequality following because $b^{\prime}$ and $\theta^{\prime}$
are both positive and because $b^{\prime}=\mu<\overline{y}$. Notice
also that the inequality is strict because we must have at least one
observation for which $z_{i}\neq0$;
otherwise, our full rank assumption would be violated, and we would
be in the case of perfect collinearity.
Because this expression is always positive, $l(\beta+k\gamma^{*})>l(\beta)$
for any $k>0$ and for any $\beta\in\mathbb{R}^{M}$.
Thus, there is no finite solution $\beta^{MLE}\in\mathbb{R}^{M}$ that maximizes $l\left(\cdot\right)$, and estimates are said not to exist.
Intuitively, the objective function will always be increasing as either $x_{i}\beta\rightarrow-\infty$ for at least one observation where $y_{i}=0$ or $x_{i}\beta\rightarrow\infty$
for at least one observation where $y_{i}=\overline{y}$.

Alternatively, suppose that, for any $\gamma$, we always have that
$x_{i}\gamma\neq0$ for at least one interior observation ($0<y_{i}<\overline{y}$).
Importantly, this ensures that $\lim_{k\rightarrow\infty}l_{i}(\beta+k\gamma)=-\infty$
for at least one observation. Since $l_{i}(\cdot)$ is continuous
in $\beta$ and (by assumption) has a finite upper bound, we can therefore
always identify a finite scalar $\overline{k}$ such that $k>\overline{k}$
implies that $l(\beta+k\gamma)=\sum_{i}l_{i}(\beta+k\gamma)<\sum_{i}l_{i}(\beta)=l(\beta)$,
for any $\beta,\gamma\in\mathbb{R}^{M}$. In other words, searching
for an ML or PML solution $\beta^{MLE}$ in any direction
from any starting point in $\mathbb{R}^{M}$ space will always eventually
yield a decrease in $l\left(\cdot\right)$. Because
$l\left(\cdot\right)$ is continuous, this guarantees the existence
of a finite $\beta^{MLE}\in\mathbb{R}^{M}$ maximizing $l\left(\cdot\right)$.

Next, note that, for any $y_{i}=0$ observation such that $x_{i}\gamma>0$
, $l_{i}(\beta+k\gamma)$ is monotonic in $k$ with $\lim_{k\rightarrow\infty}l_{i}(\beta+k\gamma)=-\infty$.
Similarly, note that $\mu<\overline{y}$ ensures the same is true
for any $y_{i}=\overline{y}$ observation such that $x_{i}\gamma<0$.\footnote{To be clear, if $\overline{y}=\infty$, we never have that $y_{i}=\overline{y};$
only conditions \eqref{eq:inner} and \eqref{eq:right} are salient.
On the other end of the spectrum, models for ``fractional'' data
such as Bernoulli PML (cf., \citealt{papke_econometric_1996,silva2014estimating})
allow the dependent variable to vary continuously over $[0,1]$. For
these models, all three conditions stated in Proposition \ref{Prop1}
are relevant.} Thus, we can again always find a sufficient $\overline{k}$ such
that $k>\overline{k}$ implies $l(\beta+k\gamma)<l(\beta)$ so long
as we always have that either $x_{i}\gamma>0$ for at least one observation
where $y_{i}=0$ or $x_{i}\gamma<0$ for at least one observation
where $y_{i}=\overline{y}$.\footnote{Readers should be wary of the weight carried by the word ``always''
here. It could be the case, for example, that $x_{i}\gamma=0$ for
all $y_{i}>0$ with $x_{i}\gamma\geq0$ for all $y_{i}=0$. This is
still a case where estimates do not exist, since $\gamma^{*}=-\gamma$
would satisfy the needed conditions.} Finally, note that we do not consider the case where there exists
a vector $\gamma$ such that $x_{i}\gamma=0$ for \emph{all $i$},
as this is the case where $X$ is not of full rank. Therefore, the
only possible scenario in which estimates do not exist is the one
where we can find a linear combination of regressors $z_{i}=x_{i}\gamma^{*}$ satisfying
\eqref{eq:inner}-\eqref{eq:left}. \hfill{}$\blacksquare$\medskip{}

To tie in some standard terminology from the binary outcome literature
(cf., \citealp{albert_existence_1984}), we will say that when estimates
maximizing \eqref{eq:glm} do not exist, the linear combination of
regressors defined by $z_{i}=x_{i}\gamma^{*}$ ``separates'' the
observations for which $z_{i}\gtrless0$ from the rest of the sample.
For the sake of providing a more unified perspective,
we will henceforth refer to the nonexistence with the term ``separation''.
A particular point of interest for us is how to also adapt the related
terms ``complete separation'' and ``quasi-complete separation''
to this more general context. For binary outcome models, separation
is usually considered ``complete'' if either $z_{i}<0$ for all
$y_{i}=0$ or $z_{i}>0$ for all $y_{i}=\overline{y}=1$, since in these
cases the value of $z_{i}$
perfectly predicts whether $y_{i}$ is $0$ or $1$. Otherwise,
we have only ``quasi-complete separation'', where only some $y_{i}$
outcomes are perfectly predicted. Outside of binary outcome models, however,
as long as $y_{i}$ takes on at least two positive
values, it will never be the case that $z_{i}=0$ perfectly predicts
all positive $y_{i}$, regardless of whether $z_{i}<0$ perfectly
predicts all $y_{i}=0$ outcomes or only some of them. Thus, for lack
of an analogous vocabulary for discussing separation in the nonbinary
outcome case, we would suggest that separation occurring in these models
should generally be regarded as ``quasi-complete''.

In addition, for those readers more familiar with \citet{santos_silva_existence_2010}'s
results for Poisson models specifically, another term that is useful
for us to clarify for the nonbinary outcome context is ``overlap''.
In \citet{santos_silva_existence_2010}, Poisson estimates are shown
to exist so long as there are no regressors that are perfectly collinear
over the subsample where $y_{i}>0$. In our way of phrasing the issue,
this criterion equates to saying there exists no linear combination
of regressors satisfying equation \eqref{eq:inner}. However, as \citet{santos_silva_existence_2010}
are careful to note, this criterion is only sufficient, rather than
necessary and sufficient. As the remaining elements of the preceding
proof show, even if such a linear combination exists, separation is
still avoided if $z_{i}$ takes on both positive and negative values
when $y_{i}=0$, such that its maximum and minimum values over $y_{i}=0$
``overlap'' the $z_{i}=0$ values it takes on when $y_{i}>0$.\footnote{The question of when overlap occurs is precisely the point left ambiguous
in \citet{santos_silva_existence_2010}. See page 311 of their paper.}

Interestingly, we do not know of a widely accepted label for what
we have called the ``linear combination of regressors that separates
the data'' (i.e., $z_{i}=x_{i}\gamma^{*}$). Clearly, $z$
plays a central role in the analysis of separation, and the literature
could use a concise name for it.
We propose the term ``certificate of separation.'' The idea is that we can
easily ``certify'' whether any such $z$ separates the data by verifying that i) its values satisfy \eqref{eq:inner}-\eqref{eq:left}, and ii) that the $R^{2}$ of a regression of $z$ against the regressors $x$ is equal to one.\footnote{Our proposed name borrows from optimization, where phrases such as
``certificate of feasibility'', ``certificate of convexity'',
``certificate of nonnegativity'', and so on are used with a similar
purpose.}
Note that there can be multiple $z$'s certifying separation of different observations, and that adding up two or more $z$'s preserves their properties.\footnote{If $z^{*}$ and $z^{**}$ are valid certificates of separation with associated coefficients $\gamma^{*}$ and $\gamma^{**}$, then $z^{***}=z^{*}+z^{{**}}$ is also a valid certificate of separation, as \eqref{eq:inner}-\eqref{eq:left} hold trivially and $z^{***}$ is a linear combination of $x$ with associated coefficient $\gamma^{{***}}=\gamma^{*}+\gamma^{**}$.} Thus, we refer to an ``overall certificate
of separation'' $\overline{z}$ that can be used to identify all
separated observations. For any $\gamma^{*}$ associated with a certificate
of separation, we will tend to use the term ``separating vector''
(although another name for it is the ``direction of recession'';
see \citealp{geyer2009likelihood}). We will use $\overline{\gamma}$
to denote the separating vector associated with $\overline{z}$.\medskip{}

\textbf{Results for gamma PML and inverse Gaussian} \textbf{PML}.
One stipulation that sticks out in Proposition \ref{Prop1} is our
requirement that the individual likelihood contribution
$l_{i}(\cdot)$ have a finite upper bound. To our knowledge, the implications
of relaxing this assumption have not been touched upon in the prior
literature. Rewinding some of the last few details behind the above
proof, the specific role played by this restriction is that it ensures
that if $\lim_{k\rightarrow\infty}l_{i}(\beta+k\gamma)=-\infty$
for any $i$, the overall objective function $l(\beta+k\gamma)=\sum_{i}l_{i}(\beta+k\gamma)$
also heads toward $-\infty$ for large $k$. However, this might not hold if $l_{i}(\cdot)$
is not bounded from above. In
this case, even if the data exhibit ``overlap'' (as defined above),
this alone will not be sufficient to ensure that $l(\cdot)$ has a
maximum. Instead, stronger conditions may be needed.

For illustration, the two models we will consider where $l_{i}(\cdot)$
does not necessarily have a finite upper bound are gamma PML and inverse
Gaussian PML.\footnote{Note that ML estimation of either a gamma distribution or an inverse
Gaussian distribution will not admit $y_{i}=0$ values. Thus, we consider
PML versions of these estimators only. In general, what gamma PML and
inverse Gaussian PML have in common is that their score functions
place a relatively larger weight on observations with a smaller conditional
mean. Similar results will apply to other estimators with
comparable score functions. } As shown in Table \ref{tab:Key-results-for}, the form of the pseudo-likelihood function  for gamma PML regression is
\begin{equation}
l(\beta)=\sum_{i}l_{i}(\beta)=\sum_{i}-\alpha\,y_{i}\exp\left(-x_{i}\beta\right)-\alpha\,x_{i}\beta\label{eq:gamma}
\end{equation}
and the form for inverse Gaussian PML is
\begin{equation}
l(\beta)=\sum_{i}l_{i}(\beta)=\sum_{i}-\alpha\frac{y_{i}}{2}\exp\left(-2x_{i}\beta\right)+\alpha\exp\left(-x_{i}\beta\right).\label{eq:Inv-Gaussian}
\end{equation}
In either case, notice that the associated $b_{i}$ function from
equation \eqref{eq:glm}, $x_{i}\beta$ for gamma and $-\exp(-x_{i}\beta)$
for inverse Gaussian, has a lower bound of $-\infty$, as $\lim_{x_{i}\beta\rightarrow-\infty}b_{i}=-\infty$.
Thus, in either case, while $l_{i}(\cdot)$ continues to have a finite
upper bound for observations where $y_{i}>0$, if $y_{i}=0$, we have
that $\lim_{x_{i}\beta\rightarrow-\infty}l_{i}(\cdot)=\infty$. With
this in mind, the following Proposition collects results that apply
to either of these estimators:

\begin{proposition} (Gamma PML and inverse Gaussian PML) Suppose
the matrix of regressors \Regressors is of full column rank. Also
let $\gamma^{*}=(\gamma_{1}^{*},\gamma_{2}^{*},\ldots,\gamma_{M}^{*})\in\mathbb{R}^{M}$
be a nonzero vector of the same dimension as $\beta$.
\begin{enumerate}
\item[(a)] If $l(\beta)$ conforms to gamma PML as stated by \eqref{eq:gamma}, PML
estimation of $\beta$ will not have a solution if and only if there
exists a linear combination of regressors $z_{i}=x_{i}\gamma^{*}$
such that
\begin{equation}
z_{i}\ge0\quad\forall\,i\quad\text{s.t.}\quad y_{i}>0
\label{eq:inner-gamma}
\end{equation}
and either of the following two conditions holds:
\begin{equation}
\sum_{i}z_{i}<0\quad\text{or}\quad\sum_{i}z_{i}=0\;\text{with \ensuremath{z_{i}>0} for at least one observation with \ensuremath{y_{i}>0}}.\label{eq:left-gamma}
\end{equation}
In addition, if only \eqref{eq:inner-gamma} can
be satisfied and $z_{i}=0$ for all $\ensuremath{y_{i}>0}$, PML estimates
of $\beta$ exist but are nonunique.
\item[(b)] If $l(\beta)$ conforms to  inverse Gaussian PML (i.e., \eqref{eq:Inv-Gaussian}),
PML estimation of $\beta$ will have no solution if and only if there
exists a linear combination of regressors $z_{i}=x_{i}\gamma^{*}$
such that $z_{i}$ satisfies \eqref{eq:inner-gamma} and at least
1 $z_{i}$ is $<0$ when $y_{i}=0$. \label{Prop2}
\end{enumerate}
\end{proposition}

Part (a) of Proposition \ref{Prop2} follows from again considering
the function $l(\beta+k\gamma)$, this time specifically for gamma
PML. Using \eqref{eq:gamma}, it is straightforward to show that $\lim_{k\rightarrow\infty}l(\beta+k\gamma)=-\infty$
if $x_{i}\gamma<0$ for at least one observation with $y_{i}>0$.
By a continuity argument similar to the one used above, this implies
that $l(\beta+k\gamma)$ must eventually become decreasing in $k$
for sufficiently large $k$.

Next, consider what happens if there exists a linear combination of
regressors $z_{i}=x_{i}\gamma$, which is always $\ge0$ when $y_{i}>0$.
In this case, because $\lim_{k\rightarrow\infty}\sum_{z_{i}\neq0}-\alpha y_{i}\exp(-x_{i}\beta-kz_{i})=0$,
we have that
\[
\lim_{k\rightarrow\infty}l(\beta+k\gamma)=\lim_{k\rightarrow\infty}\sum_{i}-\alpha\left(x_{i}\beta-kz_{i}\right)+\sum_{z_{i}=0}-\alpha y_{i}\exp(-x_{i}\beta).
\]
There are four possibilities for the above limit. If $\sum_{i}z_{i}<0$,
the gamma pseudo-likelihood function is always increasing
in the direction associated with $\gamma$, such that finite estimates
do not exist. Alternatively, if $\sum_{i}z_{i}>0$, the limit equals
$-\infty$ and we are again assured that this function must eventually
decrease with $k$, such that estimates will exist.

The remaining two possibilities occur when $\sum_{i}z_{i}=0$. In
this case, the effect of an increase in $k$ on the likelihood
function is always given by
\[
\frac{dl(\beta+k\gamma)}{dk}=\sum_{z_{i}>0}\alpha y_{i}\exp(-x_{i}\beta-kz_{i})z_{i}\ge0.
\]
Inspecting the above expression, $dl(\beta+k\gamma)/dk>0$ with strict
inequality if $z_{i}>0$ for at least one observation with $y_{i}>0$,
ensuring again that finite estimates do not exist. The final possibility
is if $z_{i}=0$ for all $y_{i}>0$ observations, in which case $l(\beta+k\gamma)=l(\beta)$
for any $k>0$. In other words, regardless of which initial $\beta$
we consider, the likelihood will always be
weakly higher when we increment $\beta$ by some positive multiple
of $\gamma$, implying either that a finite solution for $\beta$
maximizing $l(\cdot)$ does not exist (if $\sum_{i}z_{i}=0$
with $z_{i}>0$ for at least one $y_{i}>0$) or that any finite solution
will be nonunique (if $z_{i}=0$ for all $y_{i}>0$). Thus, taking
all of these results together, gamma PML estimation of $\beta$ will
not have a finite solution if there exists a linear combination of
regressors satisfying \eqref{eq:inner-gamma} and \eqref{eq:left-gamma}
and may not necessarily have a unique solution even if these conditions
are not met.

For proving part (b), which pertains instead to inverse Gaussian PML,
it is again convenient to work with the derivative of the $l(\beta+k\gamma)$
function with respect to $k$. Continuing to let $z_{i}=x_{i}\gamma$,
and after dividing up terms appropriately, this derivative can be
expressed as
\begin{align}
\frac{dl(\beta+k\gamma)}{dk} & =-\alpha\sum_{y_{i}=0}\exp\left(-x_{i}\beta-kz_{i}\right)z_{i}+\alpha\sum_{y_{i}>0}y_{i}\exp\left(-2x_{i}\beta-2kz_{i}\right)z_{i}-\alpha\sum_{y_{i}>0}\exp\left(-x_{i}\beta-kz_{i}\right)z_{i}.\label{eq:IG-diff}
\end{align}
Let us start with the conditions highlighted in part (b), where $z_{i}\ge0$
for all $y_{i}>0$ and where $z_{i}<0$ for at least one observation
where $y_{i}=0$. We can see that the second and third terms in \eqref{eq:IG-diff}
will go to $0$ in the limit where $k$ becomes infinitely large.
The first term, meanwhile, heads to infinity. Thus, the pseudo-likelihood
function increases asymptotically for large $k$, and it is clear there
is no finite solution for $\beta$.

However, we still need to verify what happens if we cannot find a
linear combination $z_{i}$ satisfying both of the conditions stated
in part (b). This part requires slightly more work. If $z_{i}\ge0$
for all $i$, for example, all three terms in \eqref{eq:IG-diff}
go to zero for $k\rightarrow\infty$\textemdash a result that is not
in itself all that informative. Likewise, if we consider what happens
when $z_{i}$ may be less than zero for $y_{i}>0$, the first and
third terms could potentially head toward $+\infty$, while the second
term heads toward $-\infty$. In all of these seemingly ambiguous
scenarios, we can use L'H\^{o}pital's rule to clarify that $dl(\beta+k\gamma)/dk<0$
for sufficiently large $k$, indicating that the pseudo-likelihood
function will always eventually decrease in the direction associated with
$\gamma.$ \hfill{}$\blacksquare$\medskip{}

To our knowledge, we are the first to study the general circumstances
under which estimates from gamma PML and inverse Gaussian PML exist.\footnote{Even \citet{wedderburn_existence_1976}, in his original derivation
of a sufficient condition for the existence of GLM estimates, specifically
avoids commenting on what conditions would be needed for gamma estimates
to exist if the dependent variable is allowed to be zero.} That these estimators have not been specifically looked at in this
context is perhaps not all that surprising, since these models have
not traditionally been used with zeroes and since the increase in
popularity of PML estimation in applied work has only occurred relatively
recently. Indeed, thanks to contributions such as \citet{manning_estimating_2001},
\citet{santos_silva_log_2006}, and \citet{head_gravity_2014}, the
main context in which researchers will likely be familiar with gamma
PML is in settings where zeroes are common, such as data for international
trade flows and health care costs. Inverse Gaussian PML is also sometimes
considered for these types of applications (see \citealp{egger_glm_2015})
but is significantly less popular, likely because it is more difficult
to work with numerically.

In this light, the results contained in Proposition \ref{Prop2} can
be read in one of two ways. On the one hand, we confirm that gamma PML
and inverse Gaussian PML can, in principle, be used with datasets
that include observed zeroes, even though their ML equivalents cannot.
Since the ability to admit zeroes on the dependent variable is one of the
reasons researchers have recently become curious about these estimators,
this confirmation seems useful.\footnote{The other main reason is that the traditional practice of applying a log transformation to the dependent variable and estimating a linear model is now widely
known to introduce a bias whenever the error term is heteroskedastic.} On the other hand, we can see from a comparison of Propositions \ref{Prop1}
and \ref{Prop2} that the criteria required for gamma PML and inverse
Gaussian PML to have finite solutions are considerably more strict
than the equivalent criteria required for most other standard GLM estimators.
Furthermore, as we will see in the next section, these fundamental
differences also imply that gamma PML and inverse Gaussian PML lack
some appealing properties that enable us to more easily remedy situations
where estimates do not exist for other models. For
these reasons, we recommend researchers to exercise extra caution
when using either of these two estimators with datasets that include
zeroes in the dependent variable.

\section{Addressing separation in practice} \label{section_solution}

This section describes recommendations for dealing with separation
in practice, including in high-dimensional environments with many
fixed effects and other nuisance parameters. Before digging into these
details, it is important to make two general points. First, as we
have shown, the implications of separation differ depending on the
estimator; thus, the appropriate remedy should similarly
depend on the estimator being used. Second, the
appeal of our own preferred alternative\textemdash {withholding the separated
observations from the estimation sample beforehand}\textemdash is likely to depend on one's comfort
level with allowing the linear predictor $x_{i}\beta$ to attain what
would ordinarily be an inadmissible value.
One method we caution against is simply removing from the model one of the regressors involved in the separation,
as this affects the identification and
estimation of all remaining parameters, with the effect differing
depending on which regressor is dropped.

In subsection~\ref{effects_of_dropping}, we will first show that when $x_{i}\beta$
is allowed to attain $\pm\infty$, separated observations often do
not affect the score function for $\beta$ under fairly general circumstances.
This insight provides a theoretical justification for the practice of {withholding} separated observations from the estimation, at which point the separation problem becomes one of perfect collinearity within the remaining estimation sample. This insight is particularly useful for models with many fixed effects, as perfect collinearity amongst the fixed effects is generally not a problem for identifying the coefficients of the non-fixed effect covariates.\footnote{{Indeed, one of the computational advantages of modern software packages for estimating models with high-dimensional fixed effects (e.g., \citealp{correia_linear_2017}) that they do not explicitly estimate unique coefficients for each fixed effect parameter. This usefully means the researcher does not need to concern themselves with which fixed effects may be collinear when implementing the estimation.}
}
Once these results are established, subsections~\ref{detecting-separation} and \ref{addressing-sep-hdfe} then focus on detecting and addressing separation, including in high-dimensional environments.

\subsection{Effects of withholding separated observations}\label{effects_of_dropping}

We now turn to discussing how identification of at least some of the
model parameters can be achieved when separation occurs. We start
with the concept utilized in \citet{aickin1979existence}, \citet{verbeek1989compactification}, \citet{geyer1990likelihood},
and \citet{clarkson_computing_1991} of a ``compactified'' (or ``extended'')
GLM where the parameter space is extended to admit its boundary values.
We can phrase this compactification in one of several equivalent ways.
For example, we could express the domain for $\beta$ as $[-\infty,+\infty]^{M}$,
the compact closure of $\mathbb{R}^{M}$.\footnote{As discussed in \citet{verbeek1989compactification}, one way to justify
the inclusion of infinitely large values in the admissible parameter
space is to observe that we could just as easily perform the maximization
over a homeomorphic space where the parameters of interest are instead
bounded by a finite interval (e.g., $[-1,1]^{M}$ instead of $[-\infty,\infty]^{M}$).
A version of this concept is also described in \citet{haberman_analysis_1974}.
It is also sometimes referred to as the ``Barndorff-Nielsen completion''
(\citealp{barndorff1978information}).} However, it is also often convenient to work with the linear predictor
$x_{i}\beta$, which in turn also may vary over $[-\infty,+\infty]$
for each $i$. In particular, note how the conditional mean $\mu_{i}$
behaves as $x_{i}\beta$ attains either of its two limits: when $x_{i}\beta\rightarrow-\infty$,
we have that $\mu_{i}\rightarrow0$, whereas when $x_{i}\beta\rightarrow\infty$
(a situation that is only relevant for binary response models and fractional
data models), we have that $\mu_{i}\rightarrow\overline{y}.$ {
It is straightforward to show that estimates for $\beta$ maximizing
the likelihood always exist when we compactify the model in this way.}

With this adjustment to the parameter space in mind, consider what
happens to the score function $s(\beta)$ and information matrix $\mathbf{F}(\beta):=\mathbb{E}[\partial s(\beta)/\partial\beta]$
in the limit as $k\rightarrow\infty$ in the case of separation outlined
above. In other words, consider
\begin{equation}
\lim_{k\rightarrow\infty}s\left(\beta+k\gamma^{*}\right):=\lim_{k\rightarrow\infty}\sum_{i}s_{i}\left(\beta+k\gamma^{*}\right):=\lim_{k\rightarrow\infty}\sum_{i}\alpha_{i}(\varphi)\left[y_{i}-\mu_{i}(x_{i}\beta+k\gamma^{*})\right]\theta^{\prime}(x_{i}\beta+k\gamma^{*};\cdot)x_{i}\label{eq:score-with-separation}
\end{equation}
and
\begin{align}
\lim_{k\rightarrow\infty}\mathbf{F}\left(\beta+k\gamma^{*}\right): & =\lim_{k\rightarrow\infty}\sum_{i}\mathbf{F}_{i}\left(\beta+k\gamma^{*}\right):=-\lim_{k\rightarrow\infty}\sum_{i}\alpha_{i}(\varphi)\mu_{i}^{\prime}(x_{i}\beta+k\gamma^{*})\theta^{\prime}(x_{i}\beta+k\gamma^{*};\cdot)x_{i}x_{i}^{\mathrm{T}}\label{eq:info-with-separation}
\end{align}
where we take $\gamma^{*}$ to be a vector satisfying the applicable
conditions for nonexistence. At this point, it will also be useful
to state the following lemma:

\begin{lemma} Suppose that $l(\beta)$ conforms to \eqref{eq:glm}.
If the likelihood contribution $l_{i}(\beta)$ has a finite upper bound, then:
\begin{enumerate}
\item[(a)] The respective limits of the $i$-specific score term, $s_{i}$, and
$i$-specific information term, $\mathbf{F}_{i}$, each go to $0$ as
the linear predictor $x_{i}\beta$ goes to $-\infty$ (i.e., $\lim_{x_{i}\beta\rightarrow-\infty}s_{i}=0$
and $\lim_{x_{i}\beta\rightarrow-\infty}\mathbf{F}_{i}=0$).
\item[(b)] For models where $\lim_{x_{i}\beta\rightarrow\infty}\mu_{i}=\overline{y}<\infty$
(e.g., binary outcome models), the limits of $s_{i}$ and $\mathbf{F}_{i}$
go to $0$ as $x_{i}\beta$ goes to $\infty$ as well (i.e., $\lim_{x_{i}\beta\rightarrow\infty}s_{i}=0$
and $\lim_{x_{i}\beta\rightarrow\infty}\mathbf{F}_{i}=0$).
\end{enumerate}
\label{Lemma1} \end{lemma} The utility of this lemma (which we prove
in our Appendix) is that, together with \eqref{eq:score-with-separation}
and \eqref{eq:info-with-separation}, it delivers the following proposition:
\begin{proposition}(Effects of withholding separated observations) Suppose
the assumptions stated in Proposition \ref{Prop1} continue to hold,
except we now consider a ``compactified'' GLM where the domain for
$\beta$ is $[-\infty,+\infty]^{M}$. Further, assume
the joint likelihood of any non-separated observations
satisfies the classical assumptions described
in \citet{gourieroux_pseudo_1984}. If there exists a separating
vector $\gamma^{*}\in\mathbb{R}^{M}$ meeting the conditions described
in Proposition \ref{Prop1}, then:
\begin{enumerate}
\item[(a)] A solution for $\beta\in[-\infty,+\infty]^{M}$ maximizing $l(\beta)$
always exists.
\item[(b)] ML and PML estimates for the linear predictors
($x_{i}\beta)$, canonical parameters ($\theta_{i}$), and conditional
means ($\mu_{i}$'s) of any observations not separated by $\gamma^{*}$
(i.e., those with $x_{i}\gamma^{*}=0$) are unaffected by dropping
any observations that are separated by $\gamma^{*}$ (i.e., those
with $x_{i}\gamma^{*}\neq0$).
\item[(c)] For any $m$ with $\gamma_{m}^{*}=0$, the associated individual
parameter estimate $\beta_{m}$ is unaffected by dropping any observations
with $x_{i}\gamma^{*}\neq0$.
\item[(d)] If \, $l(\beta)$ is in the linear exponential family and if the relationship
between the linear predictor $x_{i}\beta$ and the conditional mean
$\mu(x_{i}\beta)$ is correctly specified, {{} then
for any $m$ with $\gamma_{m}^{*}=0$ for all such $\gamma^{*}$,
the associated PML estimate for $\beta_{m}$ is consistently estimated,
and the asymptotic distributions for these estimates can be inferred using the subsample
of non-separated observations}.\label{Prop3}
\end{enumerate}
\end{proposition}

Part (a) follows from our proof of Proposition \ref{Prop1} (and is
also a central result from \citealp{verbeek1989compactification}).
After allowing $\beta$ to take on either $-\infty$ or $+\infty$,
we rule out the cases where estimates would otherwise be said not
to exist. Parts (b) and (c) are analogous to the insights contained
in \citet{clarkson_computing_1991}'s Theorem 2. After invoking Lemma
\ref{Lemma1}, the score function in \eqref{eq:score-with-separation}
can be rewritten as
\begin{equation}
\lim_{k\rightarrow\infty}s\left(\beta+k\gamma^{*}\right)=\sum_{x_{i}\gamma^{*}=0}s_{i}\left(\beta\right)+\lim_{k\rightarrow\infty}\sum_{x_{i}\gamma^{*}\neq0}s_{i}\left(\beta+k\gamma^{*}\right)=\sum_{x_{i}\gamma^{*}=0}s_{i}\left(\beta\right).\label{eq:score-with-separation-1}
\end{equation}
The key insight presented in \eqref{eq:score-with-separation-1} is
that the contribution of any observation with $x_{i}\gamma^{*}\neq0$
always drops out of the overall score function under these circumstances.
{} As a result, it must be the case that any $\beta$ that maximizes
$l(\beta)$ in the compactified model must also maximize $\sum_{x_{i}\gamma^{*}=0}l_{i}(\beta)$
(i.e., the likelihood associated with the observations \emph{not} separated by $x_{i}\gamma^{*}$).
Otherwise, we would have that $\sum_{x_{i}\gamma^{*}=0}s_{i}\left(\beta\right)\neq0$
and $\sum_{x_{i}\gamma^{*}\neq0}s_{i}\left(\beta\right)=0$, implying
that the joint likelihood of the non-separated
observations can be increased without affecting that of the separated
observations.

Parts (b) and (c) then follow because if $\beta=\beta^{*}$ maximizes the likelihood
of the non-separated observations $\sum_{x_{i}\gamma^{*}=0}l_{i}(\beta)$,
then any coefficient vector of the form $\beta^{*}+k\gamma^{*}$ maximizes
it as well. That is, the estimates of $\beta$ will be different with the separated observations than without them. But
the quantities $x_{i}\beta$, $\theta_{i}$, and $\mu_{i}$ will not
be affected, as stated in part (b), since $x_{i}(\beta^{*}+k\gamma^{*})=x_{i}\beta^{*}$
over the subsample where $x_{i}\gamma^{*}=0$ (and since $\theta_{i}$
and $\mu_{i}$ are functions of $x_{i}\beta)$. Consequently, for
any $m$ such that $\gamma_{m}^{*}=0$, the individual parameter estimate
$\beta_{m}^{*}+k\gamma_{m}^{*}=\beta_{m}^{*}$ is clearly the same
in either case, as stated in part (c).

{Part (d) then follows from similar arguments but relies on a more detailed explanation. In short, valid inference must account for the fact that some regressors become collinear when estimation is restricted to the non-separated observations. In our full proof of part (d), provided in our Appendix, we show that withholding the separated observations from the estimation is equivalent to estimating the finite components of a re-parameterized model that explicitly introduces the certificates of separation as regressors with infinite coefficients. A valid information matrix can then be formed using only the regressors that have finite estimates in this re-parameterized model, which importantly preserves all model coefficients that have finite estimates in the original model. Consequently, standard inference procedures can be used for these coefficients.}\hfill{}$\blacksquare$\medskip{}

For researchers encountering separation problems, the key takeaways
from Proposition \ref{Prop3} are likely to be parts (c) and (d):
even if one or more of the elements of the MLE for $\beta$ ``does
not exist'' (i.e., is $\pm\infty$), it is still often the case that
$\beta$ has some finite elements that are identified by the model's
first-order conditions and that can be consistently estimated. Specifically,
as long as separation is ``quasi-complete'' and there are at
least some observations with $x_{i}\overline{\gamma}=0$, coefficients
for regressors that do not play a role in the separation can be consistently
estimated by first {withholding} any separated observations, and then performing
the estimation over the subsample where $x_{i}\overline{\gamma}=0$.
Meanwhile, for the parameters that are estimated
to be infinite, one can often still estimate finite combinations of these
parameters and conduct standard inference on them, {as we show in our proof of part (d) in the Appendix}.\footnote{For example, in a Poisson model where $\mu=\exp\left[\beta_{0}+\beta_{1}x_{1}+\beta_{2}x_{2}+\beta_{3}x_{3}\right]$
and $z_{i}=x_{i1}+x_{i2}$ is a linear combination of $x_{1i}$ and
$x_{2i}$ that equals $0$ for all $y_{i}=0$ and is $<0$ for some
$y_{i}=0$, then $\exp\left[\beta_{0}+\beta_{1}z_{i}+\left(\beta_{2}-\beta_{1}\right)x_{2}+\beta_{3}x_{3}\right]$
is a re-parameterization of $\mu$ that presents the same information
about $\beta_{3}$. Here, we know that {the ML estimates for ${\beta}_{1}$ and ${\beta}_{2}$
will both be $+\infty$.} The {ML estimate for the combined parameter $\beta_{2}-\beta_{1}$ is finite}, however,
and, {importantly,} the re-parameterized model allows us to take into account its
covariance with $\widehat{\beta}_{3}$ in drawing inferences. Moreover, it can still be possible to construct one-sided confidence intervals
for the individual parameters $\beta_1$ and $\beta_2$ in cases like these; we provide an example of this type of inference on our \href{https://github.com/sergiocorreia/ppmlhdfe/blob/master/guides/README.md}{website}.\label{fn:25}}

The practical implications of these insights vary based on the model and estimator.
For binary choice models, if the data exhibit complete separation instead of only
quasi-complete separation, then meaningful estimation is
impossible with or without the separated observations. Furthermore,
Proposition \ref{Prop3} is of no use for estimators with potentially unbounded
(pseudo-)likelihood functions such
as  gamma PML, as in these cases the compactified model will
have infinitely many solutions when there is separation of any kind.
However, as we have discussed, the degree of separation for many other
commonly used GLMs can only be quasi-complete. A Poisson model, for
example, can always be estimated by first identifying and {withholding the} separated observations from the estimation sample. For these situations,
Proposition \ref{Prop3} lends significant theoretical justification
to this approach, especially when the researcher's focus is only on
a particular subset of regressors (as is often the case with fixed
effect models, for example).

To flesh out some additional intuition behind these results, it is
helpful to draw a connection between separation and the better
understood result of perfect collinearity between regressors.
Under perfect collinearity, there is at least one redundant regressor which, given the other regressors, conveys no additional information about the observed outcomes.\footnote{More precisely, this occurs when $x_{i}\gamma=0$ over the entire sample,
for some nonzero vector $\gamma$.} Therefore, the estimated effect of the redundant regressor could theoretically take any value
without affecting the score function or the estimates of variables
it is not collinear with. Separation is similar in that, because the
regressors implicated in $x_{i}\overline{\gamma}$ are only identifiable
from the observations where $x_{i}\overline{\gamma}\neq0$, {there must again be at least one regressor that
provides no information for estimating the coefficients of the regressors that are not
involved in separation.} The two issues are still fundamentally distinct,
since separation involves estimates of the problematic regressors
becoming infinite rather than indeterminate. In either case, however,
it is important that a researcher note that the choice of which regressor
to drop from the estimation performed by the estimation algorithm is often arbitrary and that the computed
coefficients of some of the remaining regressors (i.e., those that
are involved in either separation or perfect collinearity) may need
to be interpreted as being relative to an omitted regressor or omitted
regressors, as shown in \eqref{eq:reparam} {in our Appendix}. Indeed, we would generally advise that researchers should
be very cautious when a regressor is shown to be dropped by the statistical software they are using, regardless of the underlying cause.

{As another way of illustrating the difference between perfect collinearity and separation, consider the consequences of simply
dropping one of the problematic regressors but continuing to use the full sample. When the model exhibits separation, unlike in the case of perfect collinearity, dropping a regressor will have meaningful implications for both the fit of the model and the estimates of all of the model coefficients. Our suggested approach, by contrast, delivers the same model fit one would obtain using maximum likelihood to estimate the full model over the full sample. Since the separated observations are perfectly predicted by the model, their fitted values are obtained as a by-product of the initial step that detects which observations are separated. The fitted values for the remaining observations are then obtained simply by estimating the model on the subsample of non-separated observations. This estimation step also yields the maximum likelihood estimates for all coefficients that have finite estimates in the original model.}\footnote{{
As discussed in the Appendix, it is often possible to recover the signs of the coefficients whose estimates diverge under separation, and it is always possible to estimate certain finite combinations of these coefficients. In practice, recovering the signs of the infinite coefficient estimates requires first identifying the combinations of regressors
that separate the data and then carefully re-writing the linear part
of the model as in \eqref{eq:reparam} in our Appendix. Our \href{https://github.com/sergiocorreia/ppmlhdfe/blob/master/guides/README.md}{website} includes
examples of how to use our \texttt{ppmlhdfe}
Stata command to implement these steps. Also see footnote \ref{fn:25}
for simple descriptive example.}}

{As a final remark, another similarity between separation and perfect collinearity is
that separation is neither strictly a ``small sample'' issue nor
a ``large sample'' issue. It may be resolved by obtaining a larger
sample if the underlying reason is that there is not enough variation
in one or more of the regressors in the current sample. However, it
may also occur in large samples either because of fundamental co-dependence
between $y_{i}$ and some of the regressors (e.g., a trade embargo may always
predict zero exports) or because the number of regressors
increases with the sample size (as is typically the case with panel
data models and other fixed effects models commonly used in economics research).
This motivates our interest in clarifying the large-sample properties of the estimates that are obtained after withholding the separated observations.}

\subsection{Detecting separation with linear programming}\label{detecting-separation}

The discussion thus far has been strictly theoretical, but the practical
aspects of the separation problem are also interesting. To date, most
discussion in the existing literature about how to detect separation has focused on binary outcome
models, where the only relevant conditions governing separation are
\eqref{eq:right} and \eqref{eq:left}. However, for applications with nonbinary outcomes, there will usually be many observations
with $0<y_{i}<\overline{y}$, such that the third condition stated
in \eqref{eq:inner} becomes key. In some cases, this condition
can greatly simplify the task of detection. For instance, \citet{santos_silva_existence_2010} show that if $X$ is of full column rank over $0<y_{i}<\overline{y}$, then
equation \eqref{eq:inner} cannot be satisfied and there is
no separation. Likewise, if
the rank of $X$ over $0<y_{i}<\overline{y}$ is $M-1$, such that
there is only one $\gamma^{*}$ that satisfies equation \eqref{eq:inner},
it is generally easy to compute values for $z_{i}=x_{i}\gamma^{*}$
over the rest of the sample and check whether or not they satisfy
the other conditions for separation.

However, detecting separation
becomes much more
complicated if there are multiple linear combinations of regressors
that satisfy equation \eqref{eq:inner} (i.e., if $rank(X)<M-1$ over $0<y_{i}<\overline{y}$). Table \ref{twozs}
gives a simple example of a dataset that presents this issue. In
this instance, a check for perfectly collinear regressors over $y_{i}>0$
would quickly reveal that both $z_{1i}=x_{3i}-x_{4i}$ and $z_{2i}=x_{2i}-x_{4i}$
are always $0$ over $y_{i}>0$. The second- and third-to-last
columns of Table \ref{twozs} then show that both $z_{1i}$ and $z_{2i}$
exhibit overlap over $y_{i}=0$, suggesting that estimates should
exist. However, just by virtue of there being two such linear combinations
of regressors satisfying \eqref{eq:inner}, then any other linear
combination $z_{3i}$ of the form $z_{3i}=\alpha z_{1i}+(1-\alpha)z_{2i}$
also satisfies \eqref{eq:inner}. Thus, there are actually an \emph{infinite} number of linear
combinations of regressors one would need to check for overlap in
this manner in order to determine existence. In this particular example,
it is still possible to determine without too much effort that $z_{3i}=0.5z_{1i}+0.5z_{2i}$
separates the first observation. But for more general cases, a more
rigorous approach is needed to take into account the
many different ways the data could be separated.

\begin{table}[t]
\centering{}\bigskip{}
\bigskip{}
\bigskip{}
\bigskip{}
\bigskip{}
\bigskip{}
\bigskip{}
\bigskip{}
\bigskip{}
\bigskip{}
\bigskip{}
\bigskip{}
\caption{Separation may not be detected because of multiple redundant regressors
over $y_{i}>0$}
\vspace{0.1cm}
\begin{tabular}{>{\centering}p{3cm}|>{\centering}p{1cm}>{\centering}p{1cm}>{\centering}p{1cm}>{\centering}p{1cm}|>{\centering}p{1cm}>{\centering}p{1cm}>{\centering}p{1cm}}
\hline
$y_{i}$ & $x_{1i}$ & $x_{2i}$ & $x_{3i}$ & $x_{4i}$ & $z_{1i}$ & $z_{2i}$ & $z_{3i}$\tabularnewline
\hline
0 & 1 & -1 & 5 & 3 & 2 & -4 & -1\tabularnewline
0 & 1 & 2 & 0 & 1 & -1 & 1 & 0\tabularnewline
0 & 1 & 0 & -6 & -3 & -3 & 3 & 0\tabularnewline
0 & 1 & 0 & 0 & 0 & 0 & 0 & 0\tabularnewline
1 & 1 & 3 & 3 & 3 & 0 & 0 & 0\tabularnewline
2 & 1 & 6 & 6 & 6 & 0 & 0 & 0\tabularnewline
3 & 1 & 5 & 5 & 5 & 0 & 0 & 0\tabularnewline
4 & 1 & 7 & 7 & 7 & 0 & 0 & 0\tabularnewline
5 & 1 & 4 & 4 & 4 & 0 & 0 & 0\tabularnewline
\hline
\hline
\multicolumn{8}{>{\raggedright}p{14cm}}{{\footnotesize{}In this example, a typical (iterative) check for perfect
collinearity over $y_{i}>0$ would first reveal that $z_{1i}=x_{3i}-x_{4i}$
is always $0$ over $y_{i}>0$ and then subsequently also find the
same for $z_{2i}=x_{2i}-x_{4i}$. Since both $z_{1i}$ and $z_{2i}$
take on positive as well as negative values over $y_{i}=0$, it would
appear the model does not suffer from separation. However, the linear
combination $z_{3i}=.5z_{1}+.5z_{2}$ only takes on values $\le0$
over $y_{i}=0$, implying separation. $x_{1}$ is an explicit
constant.}}\tabularnewline
\end{tabular}\label{twozs}
\end{table}

In light of these complexities, \citet{silvapulle_existence_1986}
and \citet{clarkson_computing_1991} have suggested using linear programming methods
to detect separation. A suitable illustration of these approaches can be expressed using the following constrained maximization problem:
\begin{align}
\begin{split}\max_{\gamma^{S}} & \sum_{y_{i}=0}\mathds1_{x_{i}\gamma^{S}<0}+\sum_{y_{i}=\overline{y}}\mathds1_{x_{i}\gamma^{S}>0}\\
\text{s.t.} & -x_{i}\gamma^{S}\ge0\,\,\text{if \ensuremath{y_{i}=0} }\\
 & \,\,\,\,\,\,x_{i}\gamma^{S}\ge0\,\,\text{if \ensuremath{y_{i}=\overline{y}} }\\
 & \,\,\,\,\,\,x_{i}\gamma^{S}=0\,\,\text{if \ensuremath{0<y_{i}<\overline{y}}. }
\end{split}
\label{eq:LP}
\end{align}
where $\mathds1_{x_{i}\gamma^{S}<0}$ and  $\mathds1_{x_{i}\gamma^{S}>0}$ respectively denote indicator functions for observations with $x_{i}\gamma^{S}<0$ and $x_{i}\gamma^{S}>0$. If a nonzero vector, $\gamma^{S}$, can be
found that solves the problem defined by \eqref{eq:LP}, then the
linear combination $x_{i}\gamma^{S}$ clearly satisfies the conditions
for separation described in Proposition \ref{Prop1}. {Furthermore,
since $\gamma^{S}$ must maximize the number of separated observations,
it follows that $\gamma^{S}=\overline{\gamma}$.} A simplex solver
or a variety of other similar linear programming methods may be used
to solve for $\gamma^{S}$; see \citet{konis2007linear} for a thorough
discussion.

A common weakness of linear programming methods in this context is
that they suffer from the curse of dimensionality. Notice that the number
of constraints associated with \eqref{eq:LP} is equal to the number
of observations, $N$, and the number of $\gamma$-parameters
that need to be solved for is equal to the number of regressors, $M$. While there are standard operations that may be used to reduce
the size of the problem to one with only $N-M$ constraints (cf.,
\citealp{konis2007linear}, p. 64), an obvious problem nonetheless
arises if either $M$ or $N-M$ is a large number, as is increasingly
the case in applied economics research.\footnote{As noted in the introduction, this popularity is largely driven by
the wide adoption of fixed effects Poisson PML (FE-PPML) estimation for
estimating gravity models. For example, \citet{figueiredo_industry_2015}
estimate a gravity model for patent citations with $N\thickapprox26$
million and $M\thickapprox27,000$, and \citet{larch_currency_2017}
estimate a similar model for international trade flows with $N\thickapprox880,000$
and $M\approx55,000$. However, high-dimensional fixed effects estimation
is also likely to become more attractive for other GLM estimators
aside from PPML as well; see \citet{stammann2016estimating}, \citet{stammann2017fast},
and \citet{fernandez-val_individual_2016} for some relevant innovations
that have appeared in the past few years.} In these cases, the standard approach just described necessitates
solving a high-dimensional linear programming problem, which may be
difficult to solve even using the most computationally efficient linear
programming solvers currently available.\footnote{Computationally efficient linear programming solvers typically involve inverting
an $M\times M$ basis matrix (cf., \citealp{hall2011high}), a step
we would prefer to avoid.} The following discussion, therefore, turns to the question of how to
equip researchers to deal with the separation problem in models with
many fixed effects and other nuisance parameters.

\subsection{Addressing separation in high-dimensional environments\label{addressing-sep-hdfe}}

To introduce a notion of high dimensionality, we
will now suppose the set of regressors can be partitioned into two
distinct components: a set of $P$ non-fixed effects regressors $w_{i}=w_{1i},\ldots,w_{Pi}$,
which we will treat as countable in number, and a set of $Q$ indicator variables $d_{i}=d_{1i},\ldots,d_{Qi}$, where $Q$ is allowed to be
a large number. The total number of regressors $M=P+Q$ is therefore
also large, and the combined matrix of fixed effect and
non-fixed effects regressors can be expressed as $X=\{w_{i},d_{i}\}$.
Note that this partition does not
depend on the indexing of the fixed effects, but they could easily
be subdivided into multiple levels (e.g., ``two-way'' or ``three-way'' fixed effects specifications) depending on the application.\footnote{In addition, note that the high-dimensional portion of the regressor
set need not consist of only indicator variables; the methods we describe
can also be applied to models where $d_{i}$ contains linear time
trends, fixed effects interacted with non-fixed effect variables,
and so on without loss of generality.} The number of observations, $N$, is assumed to be greater than $M$, with $N-M$ also generally treated as a large number.

Before describing our preferred method for solving this problem, we
first briefly discuss the shortcomings of other feasible methods that
might otherwise seem appealing.
{} One strategy is to reduce the dimensionality of the above linear
programming problem using the Frisch-Waugh-Lovell theorem, extending
an earlier strategy proposed by \citet{larch_currency_2017}. By projecting
out all other regressors---including fixed effects---from each regressor
over the sample of positive observations, this reduces the number
of parameters to be solved from $M$ to $P$, making the problem much
more tractable. This initial projection step can be performed quickly
even for very large $Q$ using the methods of {\citet{correia_linear_2017}.
As we discuss further in the Appendix, this approach is effective
in many settings but cannot detect separation patterns that involve
only fixed effects, since the fixed effects are purged at the outset.
While the impact on the estimates of the non-fixed effect parameters
from the latter type of separation may be benign, it may
still cause numerical issues that slow or prevent convergence of the researcher's
estimation algorithm.}

As another alternative, we could simply attempt to
compute estimates without any precautions and consider any observation
for which the conditional mean appears to be converging numerically
to either $0$ or $\overline{y}$ to be separated.\footnote{\citet{ppmlhdfe}, \citet{stammann2017fast}, and \citet{berge2018efficient} each describe
algorithms that can accommodate high-dimensional
models in a computationally efficient way. The iterative output from
these algorithms can in principle be used to detect observations whose
computed $\mu$ values are converging to inadmissible values.
As discussed in the Appendix, we have made such a
method available as an option for our {\tt ppmlhdfe} command.}
This strategy has the advantage of being model-agnostic
and simple to implement, but it is generally unreliable. As noted
by \citet{clarkson_computing_1991}, it is not guaranteed to detect
separation correctly. Moreover, leaving these observations in the
regression sample, even temporarily, may again lead to non-convergence and
is likely to slow down convergence even in the best of cases. Implementing
this method is especially challenging when the true distribution of
$\mu$ is very skewed, as it becomes very difficult to numerically
distinguish true instances of $\mu=0$ from mere small values of $\mu$.

Our algorithm, which is based on an application of weighted least squares,
does not suffer from these types of issues. It can be applied to a
very general set of estimation settings, is guaranteed
to detect separation, and is both simple to understand and fast. Moreover,
it can be implemented in any standard statistical package (without
the need for a linear programming solver).

We now turn to describing how the algorithm works for the
estimation of Poisson models and similar models with only a lower
bound. We will then explain how it may be readily applied to binomial
and multinomial models without loss of generality. To proceed, let
$u_{i}$ be an artificial regressand such that $u_{i}\le0$ when $y_{i}=0$
and $u_{i}=0$ when $y_{i}>0$. Also, let $\omega_{i}$ be a set of
regression weights, given by
\begin{align*}
\omega_{i} & =\begin{cases}
1 & \text{if \ensuremath{y_{i}=0}}\\
K & \text{if \ensuremath{y_{i}>0,}}
\end{cases}
\end{align*}
with $K$ an arbitrary positive integer. The purpose behind these
definitions is that we can choose a sufficiently large $K$ such that
a weighted regression of $u_{i}$ on $x_{i}$ can be used to detect
if the equality constraint in \eqref{eq:inner} can be satisfied by
the data. {This result is an application of what is sometimes called
the ``weighting method'' (\citealp{stewart1997weighting}).}\footnote{{It is also similar to penalty methods and barrier methods, which
have been used for decades as alternatives to simplex-based algorithms
for solving linear programming problems (\citealp{forsgren2002interior}).
The value-added of our approach is that it also takes advantage of
recent computational innovations that are specific to the estimation
of least-squares regressions and that readily accommodate models with
arbitrarily high dimensionality.}} {We clarify how this technique works using the following lemma:}

\begin{lemma}For every $\epsilon>0$, there is an integer $K>0$
such that $e_{i}$, the residual from the weighted least-squares regression
of $u_{i}$ on $x_{i}$ using $\omega_{i}$ as weights, is within
$\epsilon$ of zero ($|e_{i}|<\epsilon$) for the observations where
$y_{i}>0$. \label{Lemma2} \end{lemma}

To prove this statement, note first that the residual sum of squares
(RSS) minimized by this regression will be at most $u^{\prime}u$.
It is then useful to let $K$ equal the smallest integer that is ${>u^{\prime}u/\epsilon^{2}}$.
If the weighted least-squares residual $e_{i}$ is greater than $\epsilon$
in absolute magnitude, then that observation will contribute more
than ${K\epsilon^{2}}$ to the RSS and the RSS will be at least $K\epsilon^{2}$.
If $K>{u^{\prime}u/\epsilon^{2}}$ then RSS$>u^{\prime}u$, which
is a contradiction.\hfill{}$\blacksquare$\medskip{}

Because we can force the predicted values of $u_{i}$ from this regression
to zero for observations with $y_{i}>0$, the coefficients computed
from this regression therefore satisfy \eqref{eq:inner}. The only
remaining step is to choose $u_{i}$ so that all separated observations
have predicted values less than zero and all non-separated observations
have predicted values equal to zero. We achieve this goal via the
following algorithm:
\begin{enumerate}
\item Given a certain $\epsilon>0$, define the working regressor $u_{i}$
and regression weight $\omega_{i}$ as
\begin{align*}
u_{i} & =\begin{cases}
-1 & \text{if \ensuremath{y_{i}=0}}\\
0 & \text{if \ensuremath{y_{i}>0;}}
\end{cases} & \omega_{i} & =\begin{cases}
1 & \text{if \ensuremath{y_{i}=0}}\\
K & \text{if \ensuremath{y_{i}>0.}}
\end{cases}
\end{align*}
Observe that: (i) the regressand is either zero or negative; (ii)
$u^{\prime}u$ is equal to the number of $y_{i}=0$ observations (denoted
as $N^{(0)}$).
\item Iterate on these two steps until all residuals are smaller in absolute
magnitude than $\epsilon$ (i.e., until all $|e_{i}|<\epsilon$):
\begin{enumerate}
\item Regress $u_{i}$ against $x_{i}$ using weights $\omega_{i}$. Compute
the predicted values $\widehat{u}_{i}=x_{i}\widehat{\gamma}$ and
residuals $e_{i}=u_{i}-\widehat{u}_{i}$.
\item For observations with $y_{i}=0$, update $u_{i}=\min(\widehat{u}_{i},0),$
ensuring that the regressand remains $\le0$.\footnote{One could also update $K$ and $\epsilon$ with each iteration as
well. In theory, this would lead to exact convergence. In practice,
we would typically need to insist $\epsilon$ be no smaller than $1e-16$,
which is the machine precision of most modern 64 bit CPUs. }
\end{enumerate}
\end{enumerate}
The \emph{unweighted $R^{2}$ }of the last regression iteration is
always equal to $1.0$ when it converges (i.e., $u_{i}=\widehat{u}_{i}$
for all $i$). The following proposition establishes the convergence
properties of this algorithm and its effectiveness at detecting separation:

\begin{proposition} (Convergence to the correct solution) The above
algorithm always converges. Furthermore, if all $\widehat{u}_{i}=0$
upon convergence, there exists no nonzero vector $\gamma^{*}\in\mathbb{R}^{M}$
that solves the system defined by \eqref{eq:inner} and \eqref{eq:left}
and there is no separation. Otherwise, the observations that are found
to have $\widehat{u}_{i}<0$ are separated and all the observations
with $\widehat{u}_{i}=0$ are not separated.\label{Convergence}\end{proposition}

We provide proof of this proposition in our Appendix. The main observation
for our current purposes is that none of the above steps are significantly
encumbered by the size of the data and/or the complexity of the model.
Thanks to the recent innovations of \citet{correia_linear_2017},
weighted linear regressions with many fixed effects can be computed
in almost-linear time (as can more general high-dimensional models
using time trends or individual-specific continuous regressors).\footnote{As discussed in \citet{guimaraes_simple_2010}, this is because we
can use the Frisch-Waugh-Lovell theorem to first ``partial out''
the fixed effects, $d_{i}$, from either side of the problem via a within-transformation
operation and then regress the within-transformed residuals of $u_{i}$
on those of the non-fixed effect regressors, $w_{i}$, to obtain $e_{i}$.
\citet{correia_linear_2017} then shows how to solve the within-transformation
sub-problem in nearly linear time.} The above method can therefore be applied to virtually any estimation setting for
which \eqref{eq:inner} and \eqref{eq:left} are necessary and sufficient
conditions for existence, even when the model features many levels
of fixed effects and other high-dimensional parameters. Notably, this
includes frequency table models\textemdash the original object of
interest in \citet{haberman_analysis_1974}\textemdash which themselves
may be thought of as multi-way fixed effects models without non-fixed
effect regressors.

The above algorithm still needs a name. Its defining features are
that it iteratively uses weighted least squares in combination with
a ``linear rectifier'' function\footnote{We borrow this term from the machine learning literature, where $min(y,0)$
and $max(y,0)$ are known as linear rectifiers or ReLUs (Rectified
Linear Units). Despite their simplicity, ReLUs have played a significant
role in increasing the accuracy and popularity of deep neural networks
\citep{glorot2011deep}.} to ensure $u_{i}$ eventually converges to the overall certificate
of separation that identifies all separated observations. Thus, we
have settled on the name ``iterative rectifier'' (or IR for short).\footnote{``Iteratively Rectified Weighted Least Squares'' would have introduced
acronym ambiguity with ``Iteratively Reweighted Weighted Least Squares''.}

{Finally, it is important to clarify that our iterative rectifier algorithm readily extends to a broader class of models such as binary outcome models and censored models.
For the binary outcomes---and, more generally, for multinomial discrete-choice models---the extension amounts to a simple re-parametrization.
In particular, a logit model can be rewritten as a Poisson model (see \citeauthor{albert_existence_1984}, \citeyear{albert_existence_1984}).\footnote{\citet{albert_existence_1984} conjectured that this type of equivalence between logit and Poisson models could be used to simplify the problem of detecting separation in frequency-table models. The notes we provide in our Appendix include a proof of \citet{albert_existence_1984}'s conjecture.}
For larger problems involving binary outcome models, Appendix \ref{appendix:verifying} explains---and proves---how to use this transformation to write down equivalent Poisson models that are separated if and only if the original binary outcome models are separated.
An analogous argument applies for fractional response models where $y_{i}$ can vary continuously over $[0,\overline{y}]$.
Lastly, for censored models, as hinted by \citet{clarkson_computing_1991} and shown by \citet{koll2021}, a Type I Tobit model left-censored at zero has the same separation conditions as a Poisson model; and thus our algorithm can be applied directly to this setup.\footnote{See also the \href{https://github.com/sergiocorreia/ppmlhdfe/blob/master/guides/nonexistence_examples.md\#tobit-type-i-tobit-model}{tobit example} in our companion website.}
}
\setlength{\bibsep}{.67em}

\section{Empirical Example}\label{sec:Empirical-Example}

As an illustrative example, we work from the application of \citet{baier2019widely}.
In their main analysis, a ``two-stage'' method is used to study
heterogeneity in the effects of free trade agreements (FTA). In the
first stage, a high-dimensional vector of coefficients for each FTA
and each pair of countries is estimated using Poisson PML with additional
fixed effects. In the second stage, the estimates from the first stage
are regressed on a low-dimensional set of covariates in order to examine
sources of heterogeneity. The FTA coefficients that are estimated
in the first stage differ by the direction of trade, so that, for
example, NAFTA-US-Canada and NAFTA-Canada-US are coded as separate
indicator variables, each of which switches from 0 to 1 when the NAFTA trade
agreement goes into effect in 1994. There are 910 such indicator variables
in their data set, which also includes trade flows between 69 countries
over the years 1986 to 2006.

The issue we use for illustration arises in the first stage of \citet{baier2019widely}'s
procedure. Drawing from recommended practices in the empirical FTA
literature, a model close to the one that they estimate is
\begin{align}
y_{ijt} & =\exp\left(\xi_{it}+\zeta_{jt}+\varphi_{ij}+b_{t}I_{(i\neq j)}+\sum_{A}\sum_{(i,j)\in A}\delta_{A:(i,j)}FTA_{ijt}\right)+\varepsilon_{ijt}.\label{eq:1}
\end{align}
The dependent variable $y_{ijt}$ is bilateral trade flows, triply
indexed for origin country $i$, destination country $j$, and time
$t$. The parameters $\xi_{it}$, $\zeta_{jt}$, and $\varphi_{ij}$
therefore are respectively fixed effects for origin-time, destination-time,
and origin-destination (or ``pair''). These fixed effects alternatively
may be thought of as the coefficients of the dummy variables spanning
these dimensions, an equivalence we invoke below. The ``globalization''
coefficient $b_{t}$ measures how much international trade grows relative
to each country's domestic sales in each year. It is identifiable
in spite of the fixed effects because the data set includes ``internal
trade'' observations capturing sales made by domestic producers in
their own markets. The heterogeneous FTA coefficients being estimated
are given by $\delta_{A:(i,j)}$, where $A$ denotes a unique FTA
and $(i,j)$ serves as an index for each of the origin-destination
pairs involved in that FTA. $FTA_{ijt}$ is an indicator equal to
1 when countries $i$ and $j$ are subject to an FTA.

As \citet{baier2019widely} note, it is not possible to identify a
coefficient for the effect of Romania's 1993 FTA with the European
Free Trade Area (EFTA) countries on Iceland-Romania trade. No exports
from Iceland to Romania were recorded from the beginning of the sample
until the FTA begins in 1993. Therefore, all Iceland-Romania observations
before 1993 are separated by the following linear combination:
\[
-1\times\left(D_{ISL-ROM}-FTA_{ijt}\times D_{ISL-ROM}\right),
\]
where $D_{ISL-ROM}$ is an indicator equal to 1 for the Iceland-Romania
pair. Because the model includes pair fixed effects, $D_{ISL-ROM}$
is effectively one of the regressors. Likewise, $FTA_{ijt}\times D_{ISL-ROM}$
can be regarded as a regressor whose coefficient corresponds to the
$\delta_{A:(i,j)}$ parameter for the Iceland-Romania pair upon the
signing of the EFTA-Romania FTA. It can be easily verified that this
combination induces separation. It is equal to --1 for the $y=0$
observations for Iceland-Romania before the year 1993, satisfying
\eqref{eq:left}, and is equal to 0 otherwise, satisfying \eqref{eq:inner}.

\begin{figure}[ht]
    \centering
    \caption{Agreement-by-pair FTA estimates computed using the methods and data of \citet{baier2019widely}, sorted in ascending order}
    \label{fig1}
    \includegraphics[scale=0.5]{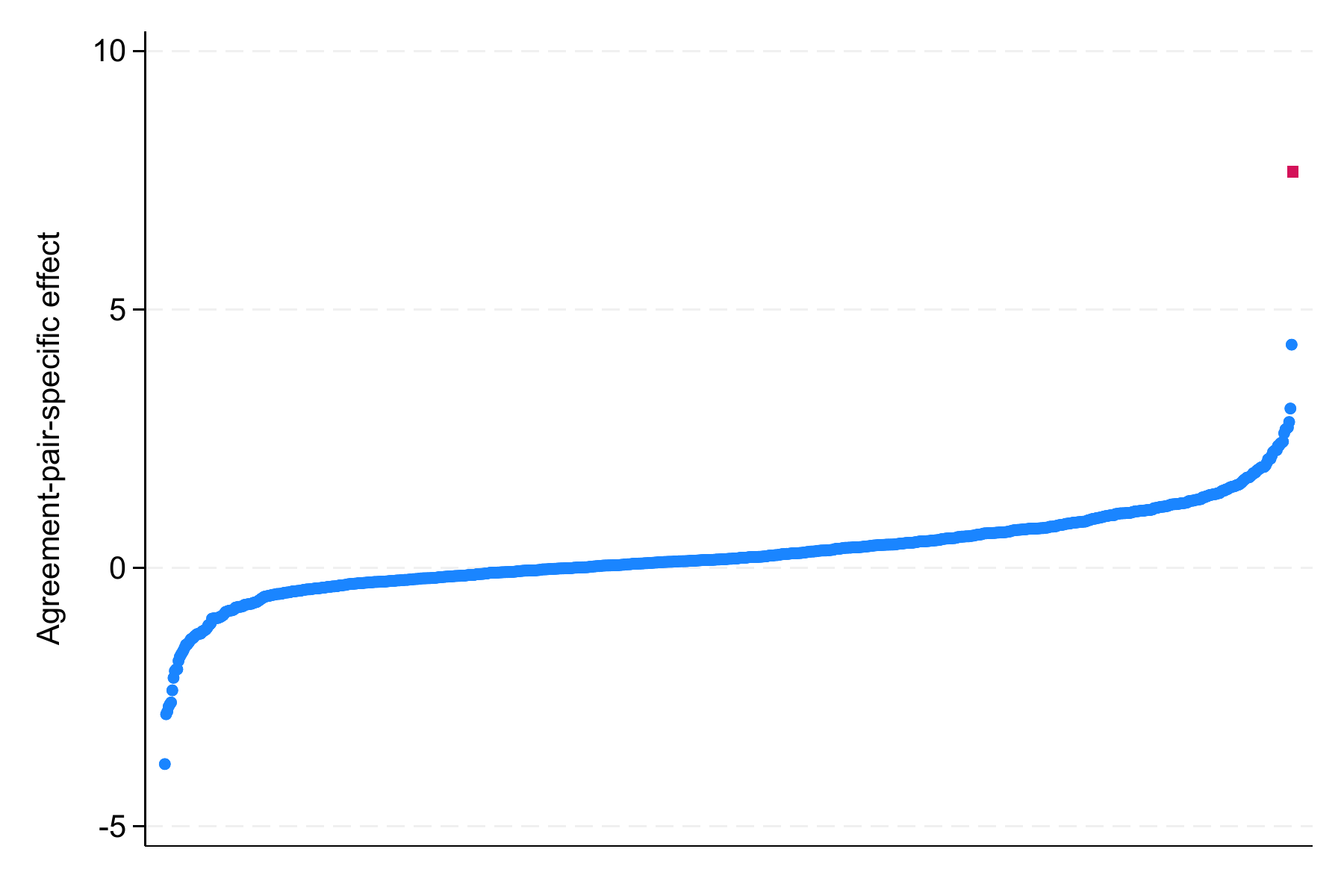}

    \begin{minipage}{\textwidth}
    \footnotesize
    The estimates are computed using the ppmlhdfe Stata command without any separation checks. The largest estimate shown (for the effect of the EFTA-Romania agreement on Iceland-Romania trade, {indicated by a red square}) is a numerical illusion. See text for more details.
    \end{minipage}
\end{figure}

Figure \ref{fig1} displays the FTA coefficient estimates we obtain
when we estimate \eqref{eq:1} using our \texttt{ppmlhdfe} command
without any checks for separation. The true coefficient estimate for
the Iceland-Romania pair should be infinity. However, though the estimated
value for Iceland-Romania is indeed the largest estimate, it does
not otherwise stand out as especially problematic given the other
extreme values that are found and the overall shape of the distribution.
Without checking for separation beforehand, a researcher could easily
mistake the reported value for the Iceland-Romania FTA estimate to
be legitimate, biasing the subsequent analysis.

This example is well chosen as a use case for our methods because
of the high degree of complexity that the model in \eqref{eq:1} embodies.
Counting all of the parameters that need to be estimated, there are
on the order of 2,800 pair fixed effects, 2,200 exporter-time and importer-time
fixed effects, 20 globalization coefficients, and 910 heterogeneous
FTA coefficients. When considering possible approaches for detecting
separation, the information matrix for this model is not straightforward
to obtain or decompose, making it difficult to implement the methods
of \citet{eck2018computationally}. Existing algorithms based on linear
programming, such as \citet{konis2007linear}, are not viable either
due to the high dimensionality.\footnote{Moreover, one can envision much larger examples along these lines. Neither the size of the data nor the number of fixed effect parameters are especially
large compared with those in, e.g., \citet{larch_currency_2017} or \citet{french2024effects}.
} When we apply our iterative rectifier
algorithm, which is implemented in \texttt{ppmlhdfe} through the \texttt{sep(ir)}
option, the 7 Iceland-Romania observations preceding their FTA are
correctly identified as being separated, as are 42 other observations
that are perfectly predicted by the pair fixed effects.\footnote{The latter 42 observations that are separated because they are associated
with pairs that never trade. Due to the pair fixed effects in the
model, all of these observations are perfectly predicted zeroes. Because
these cases each only involve a single fixed effect, they are simple
to find and can also be detected beforehand using a separate check.
For example, one can use the option \texttt{sep(fe ir)} to instruct
\texttt{ppmlhdfe} to first check for observations perfectly predicted
by any of the fixed effects and then apply the iterative rectifier
algorithm to find the remaining 7 separated observations.} The computation takes only one iteration. 

To facilitate a comparison of different methods for detecting separation,
we next reduce the original data set of \citet{baier2019widely} to
a much smaller one that retains the same structure and the same separation
issue involving Iceland-Romania trade. By randomly removing observations, while keeping Iceland-Romania, the reduced version of the data has only
1,176 observations instead of 58,989 and 14 FTA coefficients to be
estimated instead of 910. Reducing the data in this
way is helpful in part because having a smaller number of coefficients
to report allows us to verify numerically that dropping the separated
observations does not affect any of the coefficient estimates or their
standard errors when done correctly. We also remove beforehand any
pairs that never have positive trade. Table \ref{compact-example}
shows results for different approaches and options for tackling separation
applied to this reduced data set. Columns 1-3 again show results obtained
without any separation checks, only in these cases we experiment with
varying the Poisson deviance criterion used to determine if the Poisson
PML estimation algorithm has converged. In all cases, an estimate
is erroneously reported for the FTA coefficient for Iceland-Romania,
but it is interesting to observe how both the computed estimate and
its implied statistical significance depend arbitrarily on the chosen
tolerance. Column 4 then shows the results for when we apply our iterative
rectifier algorithm beforehand. As with the full data set, it performs
correctly, dropping the 7 separated observations. Furthermore, none
of the other coefficient estimates are affected by dropping these
observations as compared to column 1.

\begin{table}[htbp]
\caption{Results from applying different separation checks to reduced BYZ data}
\label{compact-example}
\footnotesize
\begin{minipage}{\textwidth}
\begin{center}
\begin{tabular}{lcccccc}
\hline
                                &\multicolumn{1}{c}{(1)}   &\multicolumn{1}{c}{(2)}   &\multicolumn{1}{c}{(3)}   &\multicolumn{1}{c}{(4)}   &\multicolumn{1}{c}{(5)}   &\multicolumn{1}{c}{(6)}   \\
\hline
EFTA-Morocco: Iceland-Morocco   &     -0.3741   &     -0.3741   &     -0.3741   &     -0.3741   &     -0.3754   &     -0.4026   \\
                                &    (0.3780)   &    (0.3780)   &    (0.3780)   &    (0.3780)   &    (0.3776)   &    (0.3679)   \\
EFTA-Romania: Iceland-Romania   &     11.3403***&      7.3403***&     16.3367   &        (omitted)       &        (omitted)       &        (omitted)       \\
                                &    (1.9027)   &    (0.8904)   &  (142.4325)   &              &            &               \\
EFTA-Romania: Romania-Iceland   &      1.8758***&      1.8758***&      1.8758***&      1.8758***&      1.8760***&      1.8770***\\
                                &    (0.1933)   &    (0.1932)   &    (0.1933)   &    (0.1933)   &    (0.1931)   &    (0.1931)   \\
Pan Arab FTA: Qatar-Kuwait      &      0.5675** &      0.5675** &      0.5675** &      0.5675** &      0.5676** &      0.5676** \\
                                &    (0.2742)   &    (0.2742)   &    (0.2742)   &    (0.2742)   &    (0.2742)   &    (0.2742)   \\
Pan Arab FTA: Egypt-Qatar       &     -0.4821** &     -0.4821** &     -0.4821** &     -0.4821** &     -0.4820** &     -0.4792** \\
                                &    (0.2337)   &    (0.2337)   &    (0.2337)   &    (0.2337)   &    (0.2338)   &    (0.2355)   \\
Agadir Agreement: Egypt-Morocco &      0.3325***&      0.3325***&      0.3325***&      0.3325***&      0.3326***&      0.3329***\\
                                &    (0.0712)   &    (0.0712)   &    (0.0712)   &    (0.0712)   &    (0.0712)   &    (0.0711)   \\
EU-EFTA: Malta-Iceland          &      2.3386***&      2.3386***&      2.3386***&      2.3386***&      2.3387***&      2.3396***\\
                                &    (0.1278)   &    (0.1278)   &    (0.1278)   &    (0.1278)   &    (0.1278)   &    (0.1274)   \\
EU-Egypt: Egypt-Greece          &     -0.4897***&     -0.4897***&     -0.4897***&     -0.4897***&     -0.4896***&     -0.4887***\\
                                &    (0.1592)   &    (0.1592)   &    (0.1592)   &    (0.1592)   &    (0.1592)   &    (0.1592)   \\
EU-Morocco: Denmark-Morocco     &     -0.7953***&     -0.7953***&     -0.7953***&     -0.7953***&     -0.7953***&     -0.7953***\\
                                &    (0.2218)   &    (0.2218)   &    (0.2218)   &    (0.2218)   &    (0.2218)   &    (0.2217)   \\
EU-Morocco: Malta-Morocco       &      1.3601***&      1.3601***&      1.3601***&      1.3601***&      1.3601***&      1.3609***\\
                                &    (0.1037)   &    (0.1037)   &    (0.1037)   &    (0.1037)   &    (0.1037)   &    (0.1038)   \\
EU-Romania: Denmark-Romania     &      0.9669***&      0.9669***&      0.9669***&      0.9669***&      0.9642***&      0.9653***\\
                                &    (0.3718)   &    (0.3718)   &    (0.3718)   &    (0.3718)   &    (0.3698)   &    (0.3699)   \\
EU-Romania: Malta-Romania       &      0.2345   &      0.2345   &      0.2345   &      0.2345   &      0.2338   &      0.2199   \\
                                &    (0.1855)   &    (0.1855)   &    (0.1855)   &    (0.1855)   &    (0.1850)   &    (0.1835)   \\
EU-Romania: Romania-Greece      &      0.6944***&      0.6944***&      0.6944***&      0.6944***&      0.6943***&      0.6950***\\
                                &    (0.2179)   &    (0.2179)   &    (0.2179)   &    (0.2179)   &    (0.2178)   &    (0.2181)   \\
EU: Malta-Greece                &     -1.0212***&     -1.0212***&     -1.0212***&     -1.0212***&     -1.0212***&     -1.0201***\\
                                &    (0.1439)   &    (0.1439)   &    (0.1439)   &    (0.1439)   &    (0.1439)   &    (0.1440)   \\
\hline
Separation check & None & None & None & IR & \texttt{ppml} & \texttt{ppml}\\
& & & & & default & ``strict'' \\
Convergence tolerance & 1e-8 & 1e-6 & 1e-10 & 1e-8 & Stata \texttt{glm} & Stata\texttt{ glm}\\
& & & & & default & default \\
Observations                               &       1,176   &       1,176   &       1,176   &       1,169   &       1,176   &       1,150   \\

\hline
\end{tabular}
\end{center}
{\footnotesize All estimates are obtained using Poisson PML estimation of \eqref{eq:1}
using a reduced version of the data set of Baier, Yotov, and Zylkin
(2019) that has only 14 FTA coefficients and 1,176 observations. Cluster-robust
standard errors, clustered by pair, are reported in parentheses. ``IR''
refers to our iterative rectifier algorithm as implemented in our
\texttt{ppmlhdfe }Stata command. See text for more details. Results
for the last two columns are obtained using the \texttt{ppml} command
of Santos Silva and Tenreyro (2010). By default, \texttt{ppml }includes
a simple check for overlap. The ``strict'' option disables this
check. The convergence criterion for \texttt{ppmlhdfe} refers to the
convergence of the Poisson deviance. The convergence criterion for \texttt{ppml}
uses Stata's default\texttt{ glm} tolerance, which is not customizable.
The correct number of separated observations is 7.}
\end{minipage}
\end{table}

The remaining columns of Table \ref{compact-example}
evaluate the performance of the methods of \citet{santos_silva_existence_2010}.
By default, their \texttt{ppml} command
checks for separation using two steps. First, it checks for collinearity
among all of the regressors over the subsample of positive observations,
which is equivalent to checking our condition \eqref{eq:inner}. Second,
if it identifies a regressor as collinear in the first step, it then
checks whether the mean value of that regressor over the positive
subsample falls between the maximum and minimum values it takes over
the sample of $y=0$ observations.\footnote{This checks a version of the relevant ``overlap''
condition in \eqref{eq:right} but differs because it focuses on whether
an individual regressor exhibits overlap rather than on linear combinations
of regressors. Importantly, finding in the first step that there is
perfect collinearity between regressors over $y>0$ often does not
indicate that any specific regressor is responsible for perfect collinearity.}
If not, that regressor is dropped. This approach
does identify the Iceland-Romania FTA indicator as problematic but
is not equipped to detect the role played by the 7 Iceland-Romania
observations preceding the FTA and thus does not drop them. Consequently,
the results in column 5 for the other coefficient estimates and their
standard errors are not the same as in the previous columns. They
are still similar in this case due to the high degree of saturation
in the model, but it is easy to appreciate how larger differences could arise
in more general settings. Finally, column 6 employs the ``strict''
separation check option for \texttt{ppml} that
only checks for collinearity over the $y>0$ sample and does not perform
the second step. In this case, \texttt{ppml} now
wrongly drops 34 observations that it takes to be perfectly predicted
by the dummy variables we use to encode the fixed effects.\footnote{\texttt{ppml} drops observations
when the excluded regressor is a dummy variable and when the less
common value of the dummy variable occurs when $y=0$. Again, this
criterion is not equivalent to our conditions \eqref{eq:inner} and
\eqref{eq:right} and thus is not able to identify the 7 observations
that should be dropped in this case.}
Again, we see numerical differences in the estimates
as well as their standard errors as a result.

Since our \texttt{ppmlhdfe} Stata
package includes several other relatively robust methods for detecting
separation, we include in our appendix an expanded version of this
example with additional results. We also include on our \href{https://github.com/sergiocorreia/ppmlhdfe/blob/master/guides/README.md}{accompanying website}
many additional examples demonstrating the implementation of our methods,
including for logit models and multinomial logit models. For example,
for logit settings, we replicate well known examples from \citet{agresti2012categorical,agresti2015foundations},
\citet{heinze2002solution}, and \citet{Kosmidis2021detectseparation}. For multinomial
logit, we replicate the ``alligators'' example from \citet{kosmidis2017multinomial}.
For Poisson and Poisson PML, we construct 17 of our own examples that
researchers working on separation detection methods can use as additional
test cases. In addition, we replicate the contingency table example
from \citet{geyer2009likelihood}, including identifying the ``direction
of recession'' in that example. We make both versions of our example
based on \citet{baier2019widely} available as well.

\section{Concluding remarks}  \label{section_conclusion}

In this paper, we have provided an updated treatment of the concept
of separation in the estimation of GLMs. While the result that all GLMs with bounded
individual likelihoods suffer from separation under similar circumstances
has been shown before by several authors, these results arguably have
not received sufficient attention. Now that estimation
techniques have progressed to the point where nonlinear models are
regularly estimated via (pseudo-)maximum likelihood with many fixed effects, there is considerable
ambiguity over whether the estimates produced by these models are
likely to exist, what it means when they do not exist, and what can
be done to ensure that the model can be successfully estimated.

We have brought more clarity to each of these topics by building on
the earlier work of \citet{verbeek1989compactification} and \citet{clarkson_computing_1991},
which we have extended to incorporate estimators that have not been previously examined and that
have their own more idiosyncratic criteria governing existence. An
important takeaway from this analysis is that some, but not all, GLM
estimators can still deliver uniquely identified, consistent estimates
of at least some of the model parameters even if other parameter estimates
are technically infinite.

We have also introduced a new method to detect separation
in models with multiple levels of high-dimensional fixed effects,
a task that would otherwise require solving
an impractical or even infeasible high-dimensional linear programming problem. As GLM estimation with
high-dimensional fixed effects increasingly becomes faster and more
appealing to researchers, the need for methods that can detect and
deal with separation in these models represents an important gap that
we aim to fill.

{At the same time, though our methods represent concrete improvements over existing practices for addressing separation in high-dimensional fixed effects settings, they are not a panacea. Fundamentally, detecting separation relies on numerically precise calculations that permit fine distinctions between observations whose predicted values truly lie on the boundary versus those whose predictions differ from it by only a small amount. Verifying these distinctions can be especially difficult for large data sets where the outcome variable is extremely skewed (e.g., sectoral trade data). In these settings, observations that are merely close to the boundary can lead to non-convergence of the estimation algorithm even if the data are not technically separated by the model. For these cases, additional safeguards such as step-halving (\citealp{marschner2011glm2}) may help to prevent erroneous infinite-valued steps before reaching convergence. If possible, it may also be useful to examine predicted values from a preliminary model to identify observations that are likely to lie near the boundary and to assess the consequences of selectively trimming some of these observations as a diagnostic exercise. Formalizing such a procedure, including the consequences for inference, may be a productive avenue for future research.}

\setstretch{1.34}
\bibliographystyle{ecta}
\bibliography{cgz_references}
\setstretch{1.35}

\appendix
\section{Appendix}

\subsection{Additional proofs}

\textbf{Proof of Lemma \ref{Lemma1}}. Recall from our discussion
of the likelihood function in \eqref{eq:glm}
that $b(\theta_{i})$ is stipulated to be increasing and convex with
respect to $\theta_{i}$. Thus, the function $l_{i}=y_{i}\theta_{i}-b(\theta_{i})+c_{i}$
has a unique, finite maximum as long as $y_{i}$ is positive. In turn,
we need only concern ourselves with how $l_{i}$ behaves either when
$y_{i}=0$ or (for part (b)) when $y_{i}=\overline{y}$. In the case
where $y_{i}=0$, note that observation $i$'s contribution to the
score function\textemdash $s_{i}(\beta)=\partial l_{i}/\partial\beta$\textemdash is
given by
\begin{align}
s_{i}(\beta) & =-\alpha_{i}(\varphi)b^{\prime}(\theta_{i})\theta^{\prime}(x_{i}\beta;\cdot)x_{i}=-\alpha_{i}(\varphi)\mu_{i}\theta^{\prime}(x_{i}\beta;\cdot)x_{i}.\label{eq:s_i_when_y_is_zero}
\end{align}
To complete the proof of part (a), first note that $\mu_{i}$ and
$\theta_{i}^{\prime}$ are both bounded from below by $0$; $\mu_{i}\times\theta_{i}^{\prime}<0$
is therefore not possible. Now suppose that $\lim_{x_{i}\beta\rightarrow-\infty}\mu_{i}\times\theta_{i}^{\prime}>0.$
In this case, the sign of $s_{i}(\beta)$ is equal to the sign of
$-x_{i}$ at the limit where $x_{i}\beta\rightarrow-\infty$. As a
result, the individual likelihood contribution $l_{i}(\cdot)$
can be perpetually increased by increasing $\beta$ in the direction
opposite to $x_{i}$ (i.e., by decreasing $x_{i}\beta$). It therefore
does not have a finite upper bound. We also need to show that $\lim_{k\rightarrow\infty}\mathbf{F}_{i}(\beta)=0$.
However, this follows directly from the fact that $\lim_{x_{i}\beta\rightarrow-\infty}s_{i}(\beta)\rightarrow0$.
To see this, let $\mathbf{F}_{i}^{(m,n)}$ denote the $m,n$th element
of $\mathbf{F}_{i}(\beta)$ and let $s_{i}^{(m)}$ denote observation
$i$'s contribution to the $m$th element of the score vector. If
observation $i$ is separated at the lower bound by some vector $\gamma^{*}$,
each element of $\mathbf{F}_{i}(\beta)$ can be written as
\[
\lim_{k\rightarrow\infty}\mathbf{F}_{i}^{(m,n)}\equiv\lim_{k\rightarrow\infty}\mathbb{E}\left[\partial s_{i}^{(m)}(\beta+k\gamma^{*})/\partial\beta_{n}\right]\equiv\lim_{k\rightarrow\infty}\mathbb{E}\left[\lim_{\tau\rightarrow0}\frac{s_{i}^{(m)}(\beta+k\gamma^{*}+\mathbf{1}_{n}\tau)-s_{i}^{(m)}(\beta+k\gamma^{*})}{\tau}\right]=0,
\]
where $\mathbf{1}_{n}$ is a unit vector with length $M$ and with
its $n$th element equal to $1$. Finally, for part (b), we also need
to consider the case where $\lim_{x_{i}\beta\rightarrow\infty}\mu_{i}\rightarrow\overline{y}<\infty$
(where we generally take $\overline{y}_{i}$ to be $1$, as is the
case in binary outcome models). A similar reasoning applies here as
well: if $\lim_{x_{i}\beta\rightarrow\infty}s_{i}(\beta)=\lim_{x_{i}\beta\rightarrow\infty}(\overline{y}-\mu_{i})\theta_{i}^{\prime}x_{i}>0$,
it will always be possible to increase $l_{i}(\cdot)$ by increasing
$\beta$ in the same direction as $x_{i}$. And we can rule out $\lim_{x_{i}\beta\rightarrow\infty}s_{i}(\beta)<0$
in these cases since $\theta_{i}^{\prime}$ cannot be negative and
since $\mu_{i}$ cannot exceed $\overline{y}$. The reasoning as to
why $\lim_{k\rightarrow\infty}\mathbf{F}_{i}(\beta)=0$ is the same
as in part (a).\hfill{}$\blacksquare$\vspace{0.5cm}

\textbf{{Proof of }}\textbf{Proposition~\ref{Prop3}(d)}.
In this proof, we establish consistency and inference for parameters
not involved in separation (i.e., where $\gamma_{m}^{*}=0$ for all
separating vectors $\gamma^{*}$), as in Proposition \ref{Prop3}(d).
To proceed, we need to establish a suitable re-parameterization of
the linear predictor $x_{i}\beta$ that preserves the same information
about any $\beta_{m}$'s associated with regressors that are not involved
in separation. Let $S<M$ denote the number of regressors for which
there exists at least one separating vector $\gamma^{*}$ with $\gamma_{m}^{*}\neq0$.
We need to allow for the possibility that there could be $J\ge1$
such separating vectors affecting the data. Without loss of generality,
we can assume the {regressors and their associated parameters} can be
ordered into the following partition:
\begin{itemize}
\item $x_{1i},\ldots,x_{Ji}$ ($J\leq S$): Regressors with parameters that \textbf{{cannot
be identified}} from the subsample of non-separated observations.
$J$ may be thought of as the number of linearly independent ``directions''
of separation.
\item $x_{J+1i},\ldots,x_{Si}$: Regressors with parameters involved in separation
($\gamma_{m}^{*}\neq0$ for some $\gamma^{*}$) that also cannot be
identified but for which we can identify certain finite \textbf{linear combinations} also involving coefficients that have infinite estimates in the original model.
\item $x_{S+1i},\ldots,x_{Mi}$: Regressors with parameters \textbf{not involved}
in separation ($\gamma_{m}^{*}=0$ for all $\gamma^{*}$). These are
the parameters referenced by part (d) of Proposition~\ref{Prop3}.
Note that valid inference for their ML estimates will still depend
on their covariances with the estimates for the other parameters of
the model.
\end{itemize}
To formalize this partition, let $\gamma^{(1)},\ldots,\gamma^{(J)}$
be $J$ linearly independent separating vectors with associated certificates
$z_{i}^{(j)}:=x_{i}\gamma^{(j)}$. That is, we can write $z_{i}^{(1)}=\sum_{m}x_{mi}\gamma_{m}^{(1)}$,
$z_{i}^{(2)}=\sum_{m}x_{mi}\gamma_{m}^{(2)}$, \ldots , $z_{i}^{(J)}=\sum_{m}x_{mi}\gamma_{m}^{(J)}$.
{Let $\Gamma:=(\gamma^{(1)}\cdots\gamma^{(J)})$ be
the $M\times J$ matrix of separating vectors and let $z_{i}=(z_{i}^{(1)},...,z_{i}^{(J)})$
be a $1\times J$ vector. Hence, $z_{i}=x_{i}\Gamma$ for all $i$.
After reordering regressors if needed, assume the first $J$ rows
of $\Gamma$ form a nonsingular submatrix }$\Gamma_{J}$. Also, continue
to let the overall certificate of separation be given by $\overline{z}:=\sum_{j=1}^{J}z^{(j)}$.
{} By Proposition~\ref{Prop1}, $z_{i}^{(j)}\leq0$ with equality for
non-separated observations, so $\overline{z}_{i}<0$ iff observation
$i$ is separated by some certificate, and $\overline{z}_{i}=0$ implies
$z_{i}^{(j)}=0$ for all $j$.

{The idea behind our re-parametrization is to replace
the first $J$ regressors with the certificates of separation. That
is, we wish to consider an otherwise equivalent model with the modified
covariate vector $\widetilde{x}_{i}:=(z_{i}^{(1)},\ldots,z_{i}^{(J)},x_{i,J+1},\ldots,x_{i,M})$.
To do so, we make use of the following equality:
\begin{align*}
\sum_{j=1}^{J}x_{ji}\beta_{j} & =\sum_{j=1}^{J}\lambda_{j}z_{i}^{(j)}-\sum_{j=1}^{J}\sum_{m=J+1}^{S}\lambda_{j}\gamma_{m}^{(j)}x_{mi},
\end{align*}
where $\lambda_{j}$ is the $j$th element of $\lambda:=\Gamma_{J}^{-1}(\beta_{1},\ldots,\beta_{J})'$
and where $\gamma_{m}^{(j)}$ is the $m$th component of $\gamma^{(j)}$.}\footnote{{This equality follows from solving the relation $z_{i}=x_{i}\Gamma$
in terms of $x_{1i},....,x_{Ji}$ and then multiplying both sides
by $(\beta_{1},\ldots,\beta_{J})'$. }}{} After making this substitution, the re-parameterized
linear index is now given by
\begin{align}
\widetilde{x}_{i}\widetilde{\beta} & :=\sum_{m=1}^{J}\lambda_{m}z_{i}^{(m)}+\sum_{m=J+1}^{S}x_{mi}\widetilde{\beta}_{m}+\sum_{m=S+1}^{M}x_{mi}\beta_{m}.\label{eq:reparam}
\end{align}
where $\widetilde{\beta}_{m}\;:=\;\beta_{m}-\sum_{j=1}^{J}\lambda_{j}\gamma_{m}^{(j)}$.
The coefficient on each of the $z_{i}^{(m)}$'s is given by the combined
term $\lambda_{m}$, while the coefficients on each of the regressors
in $x_{i,J+1},\ldots,x_{i,S}$ is now given by $\widetilde{\beta}_{m}$
instead of $\beta_{m}$, in accordance with the partition described
above. Importantly, the coefficients on the last set of coefficients
$x_{i,S+1},\ldots,x_{i,M}$ remain the same as in the original model.\footnote{{Note that for $m>S$, $\widetilde{\beta}_{m}$ reduces
to $\beta_{m}$ because $\gamma_{m}^{(j)}=0$, as these parameters
are not involved in separation. }}
Also note that, by construction, $\widetilde{x}_{i}\widetilde{\beta}\;=\;x_{i}\beta$
for all $i$.

The value of this re-parameterization is that it effectively
orthogonalizes the model with respect to the separated observations
while preserving the same likelihood. Since
by definition $z_{i}^{(1)}=z_{i}^{(2)}=...=z_{i}^{(J)}=0$ after withholding
the separated observations, it follows that {these
modified regressors provide no information about the resulting subsample
of non-separated observations. Therefore, removing these regressors
does not affect estimates or inferences obtained for the other parameters
using this subsample. The corresponding first $J$ elements of the
original coefficient vector $\beta$---i.e., $\beta_{1},...,\beta_{J}$---are
nonetheless still identifiable in the full compactified model and
should be understood to have ML estimates equal to either $+\infty$
or $-\infty$.}\footnote{{That is, the signs of the MLEs for $\beta_{1},...,\beta_{J}$
can be inferred using the definition of $\lambda$, provided the elements
of the matrix $\Gamma_{J}$ are known. Note that in the full compactified
model the estimates for each $\lambda_{m}$ coefficient are equal
to $+\infty.$ }} Estimates for $\beta_{J+1},...,\beta_{S}${{} likewise
cannot be identified without undertaking additional steps.}\footnote{In the most commonly encountered case where $J=1$, the leading coefficient $\lambda_1$ reduces to just $\beta_1/\gamma_1^{(1)}$. In this case, one can always recover the signs
of the infinite estimates for $\beta_{J+1},\ldots,\beta_S$ using simple algebra.
Our \href{https://github.com/sergiocorreia/ppmlhdfe/blob/master/guides/README.md}{website} includes
examples of how to use our \texttt{ppmlhdfe}
Stata command to recover these estimates. Interestingly, if $J\ge2$, it is not
always possible to recover estimates for $\beta_{J+1},...,\beta_{S}$
in this way. For example, if we have $J=2$, with $z_{i}^{(1)}=x_{1i}+x_{3i}$,
$z_{i}^{(2)}=x_{2i}-x_{3i}$, then it turns out the original model
coefficient for $x_{3i}$ is not identified even in the full compactified
model.}

To complete the proof, let $\widetilde{\beta}^{F}:=(\widetilde{\beta}_{J+1},\ldots,\widetilde{\beta}_{S},\beta_{S+1},\ldots,\beta_{M})$
denote the vector of finite parameters that can be estimated after
{withholding the separated observations} and let $\widehat{\beta}^{F}$
denote the corresponding vector of estimates. By parts (b) and (c)
of Proposition \ref{Prop3}, it is possible to construct a suitably
modified $(M-J)\times1$ score vector $\widetilde{s}(\widetilde{\beta}^{F}):=\sum_{\overline{z}_{i}=0}\widetilde{s}_{i}(\widetilde{\beta}^{F})$
that uniquely identifies the MLE for $\widetilde{\beta}^{F}$ using only
the subset of observations where $\overline{z}_{i}=0$. Letting $N^{\overline{z}_{i}=0}$
be the number of $\overline{z}_{i}=0$ observations, inspection of
\eqref{eq:score-with-separation-1} reveals that all elements of $\widetilde{\beta}^{F}$
will be consistently estimated if $(1/N^{\overline{z}_{i}=0})\sum_{\overline{z}_{i}=0}\widetilde{s}_{i}(\widetilde{\beta}^{F})\rightarrow_{p}0$
as $N^{\overline{z}_{i}=0}$ becomes large. Thus, consistency depends
only on the joint likelihood of the observations for which $\overline{z}_{i}=0$,
which is always well defined. As such, \citet{gourieroux_pseudo_1984}'s
proof of consistency for linear exponential families applies directly
after restricting the sample to only the non-separated observations.\footnote{Many GLMs typically used in applied economic research are from the
linear exponential family (e.g., Poisson, logit, and negative binomial).
However, a similar result can be obtained for an even more general
class of GLMs by extending the proofs of \citet{fahrmeir1985consistency}.} Similarly, a standard asymptotic variance expansion for M-estimators
gives us
\begin{align*}
\left(N^{\overline{z}_{i}=0}\right)^{\frac{1}{2}}\left(\widehat{\beta}^{F}-\widetilde{\beta}^{F}\right) & \rightarrow_{d}\mathcal{N}\left(0,\mathbf{\widetilde{F}}^{-1}\mathbf{B}\mathbf{\widetilde{F}}^{-1}\right),
\end{align*}
where $\widetilde{\mathbf{F}}:=\mathbb{E}[\partial\widetilde{s_{i}}(\widehat{\beta}^{F})/\partial\widehat{\beta}^{F}]$
is a reduced information matrix pertaining only to the finite parameters
that can be estimated and $\mathbf{B}:=\mathbb{E}[\widetilde{s_{i}}(\widehat{\beta}^{F})\widetilde{s_{i}}(\widehat{\beta}^{F})^{\mathrm{T}}]$
captures the variance of the modified score. {Since
$\widetilde{\beta}^{F}$ includes $\beta_{S+1},\ldots,\beta_{M}$,
this result clarifies how to obtain the asymptotic distribution for
these parameters as stated in the Proposition and justifies the use
of the standard sandwich variance estimator for inference. As a by-product,
it also shows that the combined parameters $\widetilde{\beta}_{J+1},\ldots,\widetilde{\beta}_{S}$
can be consistently estimated with valid asymptotic inference as well}.
\hfill{}$\blacksquare$\vspace{0.5cm}

\textbf{Proof of Proposition} \ref{Convergence}. This proof is split
into two parts. First, we show the algorithm described in Section
\ref{addressing-sep-hdfe} always converges. Then we show that if the algorithm
converges, it always converges to the correct results.\vspace{0.3cm}
 \\
 \emph{Proof of convergence}. Let $u_{i}^{(k)}$ denote the $i$th
observation of the working dependent variable at iteration $k$ and
let $\widehat{u}_{i}^{(k)}:=x_{i}\widehat{\gamma}^{(k)}$ be its predicted
value, with $\widehat{\gamma}^{(k)}$ denoting the vector of weighted
least-squares coefficients estimated at iteration $k$. Also, let $u_{i}^{2\,(k)}$
and $\widehat{u}_{i}^{2\,(k)}$ respectively denote the squares of
$u_{i}^{(k)}$ and $\widehat{u}_{i}^{(k)}$. $e_{i}:=u_{i}-\widehat{u}_{i}$
will continue to denote a residual, with $e_{i}^{(k)}$ denoting the
residual from the $k$th iteration and $e_{i}^{2\,(k)}$ denoting
its square. In addition, it will occasionally be convenient to let
$u^{(k)}$, $\widehat{u}^{(k)}$, and $e^{(k)}$ respectively denote
the vector analogues of $u_{i}^{(k)}$, $\widehat{u}_{i}^{(k)}$,
and $e_{i}^{(k)}$.

When the algorithm converges, all of the residuals from the weighted
least-squares step converge to zero: $u_{i}=\widehat{u}_{i}\le0$
$\forall i$ $\iff e_{i}=0$ $\forall i$. It would be cumbersome
to show that \emph{all} residuals indeed converge, so we instead take
a simpler route and work with the sum of squared residuals (SSR).
We can do this because $|e_{i}|\leq\sqrt{\sum_{i}{e_{i}^{2}}},$ which
implies that if the SSR converges to zero, all residuals must converge
to zero as well.

Let $SSR^{(k)}$ denote the SSR from the $k$th iteration. We will
prove that $\lim_{k\rightarrow\infty}SSR^{(k)}=0$ by first proving
that the sum given by $SSR^{(1)}+SSR^{(2)}+\ldots+SSR^{(k)}$ converges
to a finite number. To see this, note that we have by the normal equations
that $\sum{\hat{u}_{i}e_{i}}=0$. Thus, the SSR can be computed as
$\sum_{i=1}^{n}{e_{i}^{2}}=\sum_{i=1}^{n}{u_{i}^{2}}-\sum_{i=1}^{n}{\hat{u}_{i}^{2}}$
for all iterations, including $k+1$:
\[
SSR^{(k+1)}=\left(\sum_{i=1}^{n}{u_{i}^{2}}\right)^{(k+1)}-\left(\sum_{i=1}^{n}{\hat{u}_{i}^{2}}\right)^{(k+1)}.
\]
By construction, $u_{i}^{(k+1)}=\min(\hat{u}_{i}^{(k)},0)$, and thus

\[
\left(\sum_{i=1}^{n}{u_{i}^{2}}\right)^{(k+1)}=\left(\sum_{\hat{u}^{(k)}<0}{\hat{u}_{i}^{2}}\right)^{(k)}.
\]
We can also split $\sum{\hat{u}^{2,(k+1)}}$ based on the values of
$\hat{u}^{(k+1)}$:

\[
\left(\sum_{i=1}^{n}{\hat{u}_{i}^{2}}\right)^{(k+1)}=\left(\sum_{\hat{u}^{(k+1)}<0}{\hat{u}_{i}^{2}}\right)^{(k+1)}+\left(\sum_{\hat{u}^{(k+1)}\geq0}{\hat{u}_{i}^{2}}\right)^{(k+1)}.
\]
Putting the last three equations together,
\[
SSR^{(k+1)}=\left(\sum_{\hat{u}^{(k)}<0}{\hat{u}_{i}^{2}}\right)^{(k)}-\left(\sum_{\hat{u}^{(k+1)}<0}{\hat{u}_{i}^{2}}\right)^{(k+1)}-\left(\sum_{\hat{u}^{(k+1)}\geq0}{\hat{u}_{i}^{2}}\right)^{(k+1)}.
\]
If we move the last equation forward to $(k+2)$ and then add $SSR^{(k+1)}$,
we notice this is a telescoping series where one term cancels:
\[
SSR^{(k+1)}+SSR^{(k+2)}=\left(\sum_{\hat{u}^{(k)}<0}{\hat{u}_{i}^{2}}\right)^{(k)}-\left(\sum_{\hat{u}^{(k+1)}\geq0}{\hat{u}_{i}^{2}}\right)^{(k+1)}-\left(\sum_{\hat{u}^{(k+2)}<0}{\hat{u}_{i}^{2}}\right)^{(k+2)}-\left(\sum_{\hat{u}^{(k+2)}\geq0}{\hat{u}_{i}^{2}}\right)^{(k+2)}.
\]
More generally, the infinite sum of the sequence starting at $k=2$
is equal to
\[
\sum_{k=2}^{\infty}{SSR^{(k)}}=\left(\sum_{\hat{u}^{(1)}<0}{\hat{u}_{i}^{2}}\right)^{(1)}-\sum_{k=2}^{\infty}{\left(\sum_{\hat{u}^{(k)}\ge0}{\hat{u}_{i}^{2}}\right)^{(k)}}-\lim_{k\rightarrow\infty}\left(\sum_{\hat{u}^{(k)}<0}{\hat{u}_{i}^{2}}\right)^{(k)}\leq\left(\sum_{\hat{u}^{(1)}<0}{\hat{u}_{i}^{2}}\right)^{(1)}.
\]
After adding $SSR^{(1)}$ on both sides, and applying $\sum_{i=1}^{n}{u_{i}^{2}}=\sum_{i=1}^{n}{\hat{u}_{i}^{2}}+\sum_{i=1}^{n}{e_{i}^{2}}$,
we have that

\[
\sum_{k=1}^{\infty}{SSR^{(k)}}\leq\left(\sum_{\hat{u}^{(1)}<0}{\hat{u}_{i}^{2}}\right)^{(1)}+SSR^{(1)}\leq\left(\sum_{i=1}^{n}{\hat{u}_{i}^{2}}\right)^{(1)}+SSR^{(1)}=\left(\sum_{i=1}^{n}{u_{i}^{2}}\right)^{(1)}.
\]
Therefore,
\[
\sum_{k=1}^{\infty}{SSR^{(k)}}\leq\left(\sum_{i=1}^{n}{u_{i}^{2}}\right)^{(1)}=\sum_{y_{i}=0}{1},
\]
where the last equality follows from how we initialize $u_{i}$, with
$u_{i}^{(1)}=-1$ for all $y_{i}=0$ observations and $u_{i}^{(1)}=0$
otherwise. Thus, the series of SSRs is bounded above by the number of
boundary observations where $y_{i}=0$ (a finite number). We can now
show that $\lim_{k\rightarrow\infty}SSR^{(k)}=0$ (i.e., that the
sequence of SSRs converges to $0$). To see this, note that if
\[
\lim_{k\rightarrow\infty}SSR^{(k)}=c>0,\quad\text{ but}\quad\sum_{k=1}^{\infty}{\left(SSR\right)^{(k)}}\leq C
\]
for some finite $C$, then, by iteration $k^{*}:=\left\lceil \frac{C}{c}\right\rceil $,
the sum of the sequence will have exceeded $C$, a contradiction.
Therefore, the SSR converges to zero, with the same necessarily being
true for all of the individual residuals.\QED\vspace{0.3cm}
 \\
 \emph{Proof of convergence to the correct solution}. The above proof
tells us that our iterative ``rectifier'' algorithm will eventually
converge, but of course it doesn't tell us that it will converge to
the \emph{correct} solution. What we will prove now is exactly that:
\[
\lim_{k\to\infty}\hat{u}_{i}^{(k)}<0\;\text{iff the observation \textit{i} is separated}.
\]
As in the main text, $z$ is the name we will give to the ``certificates
of separation'' used to detect separated observations. By Proposition
\ref{Prop1}, any such $z$ must be a linear combination of regressors:
$z=X\gamma$, with $z_{i}=0$ if $y_{i}>0$ or if $y_{i}=0$ but is
not separated and with $z_{i}<0$ for the observations that are separated.
It is also important to keep in mind that there can be multiple $z$
vectors. We do not just want to find some of the $z$'s that induce
separation; rather, we want to identify a $z$ that is as large as
possible, in the sense of having the most nonzero rows.

Our proof that our algorithm accomplishes this task can be outlined
in two steps. We first need to show that if $\lim_{k\to\infty}\hat{u}_{i}^{(k)}<0$,
then observation $i$ is separated. This is very simple to show, since
the above proof of convergence implies that $\lim_{k\to\infty}\hat{u}_{i}^{(k)}=\lim_{k\to\infty}u^{(k)}$
and since, by construction, $u_{i}^{(k)}=0$ if $y_{i}>0$ and $u_{i}^{(k)}\le0$
if $y_{i}=0$. Recalling that $\gamma^{(k)}$ is the vector of coefficients
computed from the weighted least-squares regression in each iteration
$k$, it is now obvious that $\lim_{k\to\infty}\hat{u}^{(k)}=\lim_{k\to\infty}X\gamma^{(k)}$
is a linear combination of regressors that meets the criteria for
separation described in Proposition \ref{Prop1} if there are any
$i$ such that $\lim_{k\to\infty}\hat{u}_{i}^{(k)}<0$.

The second step, showing that $\lim_{k\to\infty}\hat{u}_{i}^{(k)}$
is necessarily $<0$ for \emph{all }separated observations, is more
complicated. To prove this part, we will rely on the following lemma:

\begin{lemma}For \uline{every possible} $z$ satisfying the criteria
for separation described in Proposition \ref{Prop1}, we must have
that $\lim_{k\rightarrow\infty}u_{i}^{(k)}<0$ on at least one row
where $z_{i}<0$.\label{LemmaA1} \end{lemma}The underlined portion
of Lemma \ref{LemmaA1} that clarifies that it applies to ``every
possible'' $z$ is important. As we will soon see, the fact that
the algorithm discovers at least one separated observation associated
with every possible linear combination of regressors that induces
separation will be sufficient to prove that $\lim_{k\to\infty}\hat{u}_{i}^{(k)}<0$
for all separated observations, completing our proof of Proposition
\ref{Convergence}. Before reaching this final step, we first need
to prove Lemma \ref{LemmaA1}:\vspace{0.3cm}
 \\
 \emph{Proof of Lemma \ref{LemmaA1}}. To prove Lemma \ref{LemmaA1},
it will first be useful to document the following preliminaries\emph{.
}First, recall that each iteration $k$ involves a regression of our
working dependent variable $u^{(k)}$ on our original regressors $X$
that produces a set of residuals $e^{(k)}$. Thus, the normal equations
for each of these regressions imply that $X'e^{(k)}=0\implies z'e^{(k)}=0$
$\forall k$ and $\forall z$ (since $z'e^{(k)}$ is just a linear
combination of $X'e^{(k)}$ that consists of premultiplying $X'e^{(k)}$
by $\gamma$). Second, we can always decompose each vector of predicted
values for our working dependent variable into its positive and negative
components using the appropriate rectifier functions: $\widehat{u}=\widehat{u}^{(+)}+\widehat{u}^{(-)}$,
where $u_{i}^{(+)}=\max(\widehat{u}_{i},0)$ and $u_{i}^{(-)}=\min(\widehat{u}_{i},0)$.
Third, using this notation, we also have that $u_{i}^{(k+1)}=\hat{u}_{i}^{(k)(-)}$
(i.e., the working dependent variable inherits the rectified predicted
values from the prior iteration).

We start with the observation noted above that the normal equations
imply $z'e^{(k)}=0$ for every iteration (i.e., $\sum{z_{i}e_{i}^{(k)}}=0$
$\forall k.$) Now let's focus on iterations $k$ and $k+1$:
\[
\sum{z_{i}e_{i}^{(k)}}+\sum{z_{i}e_{i}^{(k+1)}}=0.
\]
After grouping terms and using the definition of $e_{i}$, we have
\[
\sum{z_{i}\left[u_{i}^{(k)}-\hat{u}_{i}^{(k)}+u_{i}^{(k+1)}-\hat{u}_{i}^{(k+1)}\right]}=0.
\]
Using our decomposition, $\hat{u}_{i}^{(k)}=u_{i}^{(k),(+)}+u_{i}^{(k),(-)}$
(and likewise for $k+1$):
\[
\sum{z_{i}\left[u_{i}^{(k)}-\hat{u}_{i}^{(k),(+)}-\hat{u}_{i}^{(k),(-)}+u_{i}^{(k+1)}-\hat{u}_{i}^{(k+1),(+)}-\hat{u}_{i}^{(k+1),(-)}\right]}=0.
\]
Also recall that $u_{i}^{(k+1)}=\hat{u}_{i}^{(k),(-)}$ (and likewise
for $k+2$):
\[
\sum{z_{i}\left[u_{i}^{(k)}-\hat{u}_{i}^{(k),(+)}-u_{i}^{(k+1)}+u_{i}^{(k+1)}-\hat{u}_{i}^{(k+1),(+)}-u_{i}^{(k+2)}\right]}=0.
\]
After canceling out terms and rearranging, we have
\[
\sum{z_{i}\left[u_{i}^{(k+2)}-u_{i}^{(k)}\right]}=-\sum{z_{i}\left[\hat{u}_{i}^{(k),(+)}+\hat{u}_{i}^{(k+1),(+)}\right]}.
\]
Notice that we can focus on the observations where $z_{i}<0$ without
loss of generality, as the elements of the sum with $z_{i}=0$ are
obviously zero.\footnote{Also, there must be negative elements of $z$, as otherwise $z$ wouldn't
be a valid certificate of separation.} Thus,

\[
\sum_{z_{i}<0}{z_{i}\left[u_{i}^{(k+2)}-u_{i}^{(k)}\right]}=-\sum_{z_{i}<0}{z_{i}\left[\hat{u}_{i}^{(k),(+)}+\hat{u}_{i}^{(k+1),(+)}\right]}.
\]
Notice that unless we have reached convergence before iteration $k+2$,
then the righthand term is strictly positive. This is because $z_{i}<0$,
and $\hat{u}_{i}^{(k),(+)}$ is nonnegative, with at least one strictly
positive observation (otherwise, $\hat{u}_{i}^{(k)}$ would meet the
stopping criteria because it would be strictly nonpositive and the
next iteration would return $\hat{u}_{i}^{(k+1)}=u_{i}^{(k+1)}=\hat{u}_{i}^{(k)}.$)

Thus,
\[
\sum_{z_{i}<0}{z_{i}u_{i}^{(k+2)}}>\sum_{z_{i}<0}{z_{i}u_{i}^{(k)}}.
\]

By itself, this statement is interesting, because it tells us that
the weighted sum of $u$ is increasing as we iterate. But we can get
a useful bound if we recall that on the first iteration, $u_{i}=-1$
when $y_{i}=0,$ which implies $\sum{z_{i}u_{i}^{(1)}}=\sum{z_{i}}$.
Then, for every odd iteration with $k\geq3$, we know that

\[
\sum{z_{i}u_{i}^{(k)}}>-\sum{z_{i}}.
\]
Denote the minimum (i.e., most negative) $z_{i}$ as $z_{\min}$.
Then,
\begin{align*}
z_{\min}\sum{u_{i}^{(k)}}\geq\sum{z_{i}u_{i}^{(k)}} & >-\sum{z_{i}}\,\,\,\text{for}\,\,k=3,5,7,\ldots\\
\implies\sum_{z_{i}<0}{u_{i}^{(k)}} & <-\frac{\sum{z_{i}}}{z_{\min}}\,\,\,\text{for}\,\,k=3,5,7,\ldots
\end{align*}
Given that both $z_{\min}$ and $\sum{z_{i}}$ must be $<0$
for there to be separation, it follows that $\sum_{z_{i}<0}{u_{i}^{(k)}}$
must be negative on every odd iteration starting with $k=3$. Obviously,
this is not possible unless at least one $u_{i}^{(k)}$ is negative
for an observation where $z_{i}<0$ for each of these iterations.
The lemma follows by considering $k\rightarrow\infty$, since $u_{i}^{(k)}$
must eventually converge to the same result for both odd and even
$k$.\QED\vspace{0.3cm}
 \\
 For the remainder of the proof of the overall theorem, let $\hat{u}_{i}^{(\infty)}:=\lim_{k\to\infty}\hat{u}_{i}^{(k)}$
denote the solution obtained by our algorithm. Thanks to the insights
established by Lemma \ref{LemmaA1}, we can now prove that $\hat{u}_{i}^{(\infty)}<0$
if and only if there exists a $z$ that separates observation $i$.
We can do so by considering two cases. First, note that if $\hat{u}_{i}^{(\infty)}=0$
for all $i$, then Lemma \ref{LemmaA1} implies there cannot be any
such $z$ and the data are not separated. The more interesting case
is if $\hat{u}_{i}^{(\infty)}<0$ for at least one $i$. In that case,
recall that $\hat{u}_{i}^{(\infty)}$ is an admissible $z$ (because
it will have converged to a vector that is $<0$ for some observations
where $y_{i}=0$ and is $0$ otherwise.) Thus, for the algorithm to
fail to identify a separated observation, it would have to be the
case that there exists some other certificate of separation $z^{*}$
that is $<0$ for at least one observation where $\hat{u}_{i}^{(\infty)}=0$.
To see why this cannot happen, let
\begin{align*}
\alpha^{*} & :=\sup_{\hat{u}_{i}^{(\infty)}<0}\frac{z_{i}^{*}}{\hat{u}_{i}^{(\infty)}}>0.
\end{align*}
Then, given $z^{*}$ and our solution $\hat{u}_{i}^{(\infty)}$, we
can construct a third certificate $z^{**}$ that also separates the
data:

\[
z_{i}^{**}:=z_{i}^{*}-\alpha^{*}\hat{u}_{i}^{(\infty)}\le z_{i}^{*}-z_{i}^{*}=0,
\]
where the inequality follows from the definition of $\alpha^{*}$.
By construction, $z_{i}^{**}<0$ for at least one observation where
$\hat{u}_{i}^{(\infty)}=0$, and $z_{i}^{**}=0$ for at least one
observation where $\hat{u}_{i}^{(\infty)}<0$. If $z_{i}^{**}=0$
for \emph{all} observations where $\hat{u}_{i}^{(\infty)}=0$, we
have a contradiction, since Lemma \ref{LemmaA1}(b) tells us at least
one observation separated by $z_{i}^{**}$ must also be separated by
our solution $\hat{u}_{i}^{(\infty)}$. If not, we repeat: let
\begin{align*}
\alpha^{**} & :=\sup_{\hat{u}_{i}^{(\infty)}<0}\frac{z_{i}^{**}}{\hat{u}_{i}^{(\infty)}}
\end{align*}
and
\begin{align*}
z_{i}^{***} & =z_{i}^{**}-\alpha^{**}\hat{u}_{i}^{(\infty)},
\end{align*}
which gives us yet another certificate of separation $z_{i}^{***}$
that will equal $0$ for at least one observation where $\hat{u}_{i}^{(\infty)}<0$
and $z_{i}^{**}<0$. We can repeat this process as many times as needed
until we eventually obtain a $z$ that does not separate any observations
for which $\hat{u}_{i}^{(\infty)}<0$. Lemma \ref{LemmaA1} again
provides the needed contradiction indicating that this cannot happen.\QED\vspace{0.3cm}
 \\
 \textbf{Example code}. The number of steps needed in the above proof
may suggest the iterative rectifier algorithm is rather complicated.
However, in practice, it requires only a few lines of code to implement.
Below, we provide some generic ``pseudo code'' that should be simple
to program in virtually any statistical computing language (e.g.,
R, Stata, Matlab).\vspace{-0.3cm}

\noindent \noindent\begin{minipage}[t]{1\columnwidth}\noindent \begin{center}
\begin{tabular}{|ccc||c||c|}
\hline
\multicolumn{5}{|l|}{\emph{Pseudo code:}}\tabularnewline
 & \multicolumn{4}{l|}{Set $u_{i}=-1$ if $y_{i}=0$; $0$ otherwise}\tabularnewline
 & \multicolumn{4}{l|}{Set $\omega_{i}=K$ if $y_{i}>0$; $1$ otherwise}\tabularnewline
 & \multicolumn{4}{l|}{\textbf{Begin loop}: }\tabularnewline
 &  & \multicolumn{3}{l|}{Regress $u$ on $X$, weighting by $\omega$ (\emph{produces coefficients
$\widehat{\gamma}$})}\tabularnewline
 &  & \multicolumn{3}{l|}{Set $\widehat{u}=X\widehat{\gamma}$}\tabularnewline
 &  & \multicolumn{3}{l|}{Set $\widehat{u}=0$ if $|\widehat{u}|<\epsilon$}\tabularnewline
 &  & \multicolumn{3}{l|}{\textbf{Stop} if $\widehat{u}_{i}\le0$ for all $i$ (\emph{all separated
observations have been identified})}\tabularnewline
 &  & \multicolumn{3}{l|}{Replace $u_{i}=\min\left(\widehat{u}_{i},0\right)$}\tabularnewline
 & \multicolumn{4}{l|}{\textbf{End loop}.}\tabularnewline
\hline
\end{tabular}
\par\end{center}\end{minipage}\vspace{0.3cm}
 \\

For readers interested in more details, we have created a \href{https://github.com/sergiocorreia/ppmlhdfe/blob/master/guides/README.md}{website} that
provides sample Stata code and datasets illustrating how all of the
methods for detecting separation described in this paper can be implemented
in practice. Also see our companion paper for the \texttt{ppmlhdfe}
Stata command (\citealp{ppmlhdfe}), which provides further useful
information related to technical implementation and testing.

\subsection{An alternative method using linear programming}

\citet{larch_currency_2017} have also recently proposed a method
for detecting separation in Poisson-like models
in the presence of high-dimensional fixed effects. In their paper,
this is accomplished by first ``within-transforming'' all non-fixed
effect regressors with respect to the fixed effects, then checking
whether the within-transformed versions of these regressors satisfy
conditions for separation. As they discuss (and as we will document
here as well), any method based on this strategy is only able to detect
instances of separation that involve at least one non-fixed effect
regressor; it cannot be used to detect separation involving only the
fixed effects.
Another difference is that \citet{larch_currency_2017} describe how
to detect linear combinations of regressors that satisfy \eqref{eq:inner}
only. Detecting linear combinations of regressors that satisfy both
of the relevant conditions described in Proposition \ref{Prop1} (i.e.,
both \eqref{eq:inner} and \eqref{eq:left}) requires an appropriate
extension of their methods that incorporates the linear programming
problem in \eqref{eq:LP}.

The first step is to regress each non-fixed effect regressor $w_{pi}$
on every other regressor (including the fixed effects) over $0<y_{i}<\overline{y}$.
If we find that $w_{pi}$ is perfectly predicted over $0<y_{i}<\overline{y}$,
then we know there is a linear combination of regressors involving
$w_{pi}$ that satisfies \eqref{eq:inner}, as shown by \citet{larch_currency_2017}.
\citet{larch_currency_2017} do not discuss how this step is applicable
to the linear programming problem in \eqref{eq:LP}, but focusing
on these ``candidate'' linear combinations that we already know
to satisfy \eqref{eq:inner} turns out to be an effective way of reducing
the dimensionality of the problem (for nonbinary outcome models at
least). More formally, we can determine candidate solutions for $\gamma^{*}$
by first computing the following linear regression for each $w_{pi}$:
\begin{align}
w_{pi} & =w_{i}^{-p}\delta_{p}+d_{i}\xi_{p}+r_{pi}\,\,\,\text{for\,\,\ensuremath{0<y_{i}<\overline{y}}},\label{eq:reduce-dimensionality}
\end{align}
where $w_{i}^{-p}$ is the set of other non-fixed effect regressors
(i.e., excluding $w_{pi}$). $\delta_{p}$ and $\xi_{p}$ are the
coefficient vectors to be estimated. Our focus is on the residual
error $r_{pi}$ obtained from each of these regressions. If $r_{pi}$
is uniformly zero, then some combination of the fixed effects and
the other non-fixed effect regressors perfectly predicts $w_{pi}$
over $0<y_{i}<\overline{y}$. Or, to cement the connection with Proposition
\ref{Prop1}, we would have that $r_{pi}=w_{pi}-w_{i}^{-p}\delta_{p}-d_{i}\xi_{p}$
is a linear combination of regressors that satisfies condition \eqref{eq:inner-gamma}.

Because the estimation expressed in \eqref{eq:reduce-dimensionality}
is a linear regression, it can generally be computed very quickly
using the algorithm of \citet{correia_linear_2017}, even for models
with very large $M$. The main advantage of this first step is that
it greatly reduces the dimension of the linear programming problem
we need to solve. This is for two reasons. First, it allows us to
effectively perform a change of variables from $x_{i}$ (which is
of dimension $M$) to the set of $r_{pi}$ associated with any regressors
that are perfectly predicted over $0<y_{i}<\overline{y}$ (which will
have a much smaller dimension $\le P$$\ll M$). Second, since any
linear combination of these $z_{pi}$'s is assured to satisfy \eqref{eq:inner},
we no longer need the third set of constraints stipulated in \eqref{eq:LP}.
Since we very often have that $0<y_{i}<\overline{y}$ for a majority
of the observations in nonbinary outcome models, changing variables
in this way is likely to also greatly reduce the number of constraints.\footnote{For this reason, this first step of regressing each regressor on every
other regressor can be beneficial even in non-high-dimensional environments
when the number of observations with $0<y_{i}<\overline{y}$ is large.
A similar first step also appears in \citet{santos_silva_existence_2010}
and \citet{larch_currency_2017}, but both of these papers stop short
of verifying the ``overlap'' conditions described in \eqref{eq:right}-\eqref{eq:left}.
Addressing the latter complication requires one of the methods described
in this paper.}

A suitable reparameterization of our original linear programming
problem in \eqref{eq:LP} helps to illustrate the idea behind this
change of variables. Let $r_{i}^{*}:=\{r_{pi}|r_{pi}=0\,\,\text{if }$$0<y_{i}<\overline{y}$\},
i.e., a vector consisting of the predicted residuals from \eqref{eq:reduce-dimensionality}
associated with any $w_{pi}$ that are perfectly predicted over $0<y_{i}<\overline{y}$.
The modified linear programming problem based on $r_{i}^{*}$ instead
of $x_{i}$ is 
\begin{align}
\begin{split}\max_{\phi} & \sum_{y_{i}=0}\boldsymbol{1}\left(r_{i}^{*}\phi<0\right)+\sum_{y_{i}=\overline{y}}\boldsymbol{1}\left(r_{i}^{*}\phi>0\right)\\
\text{s.t.} & -r_{i}^{*}\phi\ge0\,\,\text{if \ensuremath{y_{i}=0}, }\\
 & \,\,\,\,\,\,\,r_{i}^{*}\phi\ge0\,\,\text{if \ensuremath{y_{i}=\overline{y}}, }
\end{split}
\label{eq:LP-HDFE}
\end{align}
where, as noted, the number of parameters we need to solve for (i.e.,
the length of the vector $\phi$ in this case) is only equal to the
number of $w_{pi}$ that we found to be perfectly predicted by other
regressors in the first step. Furthermore, the number of constraints
we need to take into account is only $N_{y_{i}=0}+N_{y=\overline{y}}$
instead of $N.$ To appreciate why this approach works, consider what
happens when a nonzero vector $\phi^{*}$ can be found solving \eqref{eq:LP-HDFE}.
In that case, $r_{i}^{*}\phi^{*}=\sum_{p|r_{pi}\in r_{i}^{*}}\phi_{p}^{*}r_{pi}=\sum_{p|r_{pi}\in r_{i}^{*}}\phi_{p}^{*}(w_{pi}-w_{i}^{-p}\delta_{p}-d_{i}\xi_{p})$
is a linear combination of regressors that satisfies \eqref{eq:inner}-\eqref{eq:left},
indicating separation.

However, while this approach is able to quickly identify separation
involving complex combinations of both fixed effect and non-fixed effect
regressors, it cannot be easily used to identify separation involving
only fixed effects (at least not without estimating \eqref{eq:reduce-dimensionality}
$M$ times in the first step, which is likely to be time consuming).
For some standard fixed effect configurations, this latter problem
is not so severe. For example, the trivial case where a fixed effect
dummy is always equal to zero when $0<y_{i}<\overline{y}$ is very
easy to find. Models with only one level of fixed effects are thus
easy to deal with in this regard. Similarly, in models with two levels
of fixed effects (e.g., exporter and importer, firm and employee),
the graph-theoretical approach of \citet{abowd2002computing} can
be applied to identify any combinations of fixed effects that are
perfectly collinear over $0<y_{i}<\overline{y}$, which then can be
added as needed to the linear programming step in \eqref{eq:LP-HDFE}.

For more general cases, such as nonbinary outcome models
with more than two levels of high-dimensional fixed effects, it has
been known since \citeauthor{haberman_analysis_1974} (1974, Appendix
B) that separation only involving categorical dummies (i.e., fixed
effects) can be difficult to verify (see also \citealp{albert_existence_1984},
p. 9.) To our knowledge, this problem has remained unresolved in the
literature and \citet{abowd2002computing}'s method cannot be used
to solve the problem for general cases either.\footnote{Abowd's method can still be used
to detect separation involving only one or two levels of fixed effects,
but not separation involving three or more levels of fixed effects, or separation also involving
the non-fixed effect regressors.
In addition, it is worth clarifying that neither perfect collinearity
between fixed effects nor separation involving only the fixed effects
(by Proposition \ref{Prop3}) poses an issue for identification of
the non-fixed effect parameters. However, separation can affect an
estimation algorithm's ability to reach convergence, the speed at
which it converges, and even whether the algorithm converges to the correct estimate values.} Thus, unless we have a nonbinary outcome model with either one or
two levels of fixed effects, we require a different method for detecting
separation.\footnote{One possible method is the one discussed in \citet{clarkson_computing_1991}
on p. 424, which allows estimation to proceed without precautions
and iteratively drops any observations that appear to be converging
to a boundary value. The algorithm we describe later in this Appendix
could be used in conjunction with this approach. However, as \citet{clarkson_computing_1991}
note, this method is not guaranteed to detect separation accurately.
Furthermore, in our own implementations, we have noted that removing
separated observations mid-estimation generally leads to slower convergence.
Yet another problem arises if ``cluster-robust'' standard errors
are used. In that case, the algorithm of \citet{correia_singletons_2015}
must also be repeatedly applied in order to determine that the correct
number of non-singletons clusters that are left as additional separated
observations are removed. Otherwise, statistical significance will
tend to be overstated.} Noting that a logit model can be transformed into a Poisson model
by adding a fixed effect (as we discuss next), the same is also true
for binary outcome models with more than one fixed effect.

\subsection{Verifying separation in binary outcome models using the logit-Poisson
transformation}\label{appendix:verifying}

While the discussion in Section \ref{addressing-sep-hdfe} focuses on the case
of a model with only a lower bound at zero, our methods can be applied
to binary response models without loss of generality. The only further
complication that is needed is that we must first transform the model
by taking advantage of the following property:

\begin{definition} (The logit-equivalent Poisson model)\textbf{ }Any
logit model with $p(y_{i}=1|x_{i})=\exp(x_{i}\beta)/[1+\exp(x_{i}\beta)]$
can be rewritten as a logit-equivalent Poisson model via the following
steps:
\begin{enumerate}
\item Let each observation now be given by $y_{i,a}$ and be indexed by
$i$ and $a=1,2$. A ``$y_{i,1}$'' will henceforth indicate an
``original'' observation from the original logit model and a ``$y_{i,2}$''
will indicate an ``artificial'' observation. The construction of
artificial observations is described in the next step.
\item For every original observation with $y_{i,1}=0$, create an artificial
observation with $y_{i,2}=1$. For every original observation with
$y_{i,1}=1$, similarly create an artificial observation with $y_{i,2}=0$.
For all artificial observations, set all corresponding elements of
$x_{i,2}$ equal to $0$. The number of observations should now be
$2N$, where $N$ is the original sample size.
\item Add a set of $i$-specific fixed effects to the model, to be given
by $\delta_{i}$. These may be thought of as the coefficients of a
set of dummy variables $d_{i}$, which equal 1 only for the two observations
indexed by a particular $i$ (one original observation and one artificial
observation).
\end{enumerate}
The resulting logit-equivalent Poisson model is given by $E[y_{i,a}|x_{i},\delta_{i}]=\exp[\delta_{i}+x_{i,a}\beta]$
and is estimated using Poisson regression.\end{definition}

After obtaining a Poisson model in this way, we have the following
equivalence:

\begin{proposition} (Logit-Poisson Equivalence) The logit-equivalent
Poisson model is equivalent to the original logit model. In particular:
\begin{itemize}
\item The first-order conditions (FOCs) for $\beta$ are the same.
\item The parameter estimates for $\beta$ and their associated asymptotic
variances are the same.
\item The conditional mean from the Poisson model equals the conditional
probability that $y_{i}=1$ from the logit model.
\end{itemize}
\label{Equivalence}\end{proposition}

The properties described in Proposition \ref{Equivalence} can be
established using the Poisson FOCs for $\delta_{i}$ and $\beta$:
\begin{align}
\sum_{i=1}^{N}\,x_{i,1}\,\left(y_{i1}-e^{\delta_{i}+x_{i,1}\beta}\right) & =0, & \forall i:\;\;\left(1-e^{\delta_{i}}\left(1+e^{x_{i,1}\beta}\right)\right) & =0,\label{eq:2}
\end{align}
where we have used the fact that $y_{i,1}+y_{i,2}=1$ and the fact
that all elements of $x_{i,2}=0$. It should be apparent that $e^{\delta_{i}}=1/(1+e^{x_{i}^{(1)}\beta})$.
After plugging in the solution for $\delta_{i}$ into the FOC for
$\beta$, we obtain
\begin{align*}
\sum_{i=1}^{N}\,x_{i,1}\,\left(y_{i,1}-\frac{e^{x_{i,1}\beta}}{1+e^{x_{i,1}\beta}}\right) & =0,
\end{align*}
which is the same as the FOC for $\beta$ from the original logit
model. The estimates for $\beta$ therefore are the same across both
models, as are the associated asymptotic variances. Furthermore, the
Poisson conditional mean $e^{\delta_{i}+x_{i,1}\beta}=\exp(x_{i}\beta)/[1+\exp(x_{i}\beta)]$
is the same as $p(y_{i}=1|x_{i})$ from the logit model.\hfill{}$\blacksquare$\vspace{0.3cm}

The most important implication of these results for our current purposes
is the following:

\begin{proposition} (Equivalence under separation) Suppose that $l(\beta)$
conforms to \eqref{eq:glm}, the matrix of regressors \Regressors
is of full column rank, and the individual log-likelihood $l_{i}(\beta)$
always has a finite upper bound. Any binary outcome model that satisfies
these conditions is separated if and only if the logit-equivalent
Poisson model is separated.\end{proposition}

Suppose we have a binary outcome model and there exists a nonzero
separating vector $\gamma^{*}\in\mathbb{R}^{M}$ that satisfies \eqref{eq:right}
and \eqref{eq:left}. Then, for any separated observation with $y_{i}=1$,
the FOC for $\delta_{i}$ in the logit-equivalent Poisson model must
satisfy $e^{\delta_{i}}=\lim_{x_{i,1}\beta\rightarrow\infty}1/(1+e^{x_{i,1}\beta})=0$
in the compactified model where such solutions are admissible. Thus,
the artificially created observation associated with $i$ ($y_{i,2}$)
has a conditional mean of $\mu_{i,2}=0$ and must be separated. Similarly,
for any separated observation with $y_{i}=0$, the conditional mean
for $y_{i,1}$, $\mu_{i,1}$, must be $0$. This can only be true
if $y_{i,1}$ is separated.

If we instead consider separation in the Poisson model, we can simply
focus on cases where either the original observation has a conditional
mean of $0$ or the artificially created observation has a conditional
mean of $0$. In the former case, it is obvious there is separation
in either model. In the latter case, we must have that $\delta_{i}=-\infty$,
which can only be true if $l(\beta)$ is increasing as $x_{i}\beta\rightarrow\infty$,
implying $y_{i,1}$ is separated in the original logit model.

Finally, the conditions for a binary outcome model to be separated
depend only on the configuration of the data and do not depend on
the specific choice of model (e.g., logit vs. probit). Therefore,
the Poisson model described above can be used to check for separation
in any conceivable GLM binary outcome model for which the individual
likelihood function is bounded from above, not just the logit model.\hfill{}$\blacksquare$

\subsection{Expanded empirical example}

Our example based on \citet{baier2019widely} in the main text focused
on the application of our iterative rectifier algorithm as we have
implemented it in our \texttt{ppmlhdfe} Stata package. Because \texttt{ppmlhdfe
}provides several other options that can potentially be used to detect
separation in high-dimensional environments, we now consider an expanded
version of this example that demonstrates and compares these options.

Table \ref{appx-table} provides a simplified set of results for a
variety of different methods and options available through \texttt{ppmlhdfe. }The
different columns respectively indicate the syntax used, the number
of separated observations detected, whether the set of observations identified
as being separated was the correct one, and whether an estimate for the
problematic regressor was erroneously reported in the subsequent regression.
The main options governing the separation checks, included within
the \texttt{sep() }wrapper are:
\begin{itemize}
\item \texttt{fe}: Drops any observations that are perfectly predicted by
a single fixed effect. This check is carried out by checking if there
are any fixed effects for which the outcome variable is always 0.
\item \texttt{simplex}: uses the augmented linear programming-based method
described in this Appendix that first partials each non-fixed effect regressor
over the subsample of positive observations.
\item \texttt{mu}: uses the method described in \citet{clarkson_computing_1991}
of removing observations during the process of estimation if their
predicted mean value appears to be converging to zero.\footnote{The \texttt{mu} method requires a tolerance parameter to determine if the current predicted value for an observation
is numerically close to zero. Our procedure first divides the $y$ variable  by its standard deviation in order to make this tolerance scale-invariant. After this standardization, we use a baseline tolerance of $1e-6$. Because some datasets are very skewed and may have very low (but positive) $y$ values, we do an extra adjustment. If $min(\log \mu | y>0)$ is below -5, we make the tolerance more conservative by that amount. For instance, if $min(\log \mu | y>0)$ = -8, then we set the tolerance for $\log \mu$ as $\log(1e-6) + (-8 - -5) = \log(1e-6) -3 = -16.82$, such that the tolerance for $\mu$ becomes $e^{-16.82} \approx 5e-8$.}
\item \texttt{ir}: our iterative rectifier algorithm, discussed in the main
text in Section \ref{addressing-sep-hdfe}.
\end{itemize}
In addition, it is possible to use an ensemble of these methods. For
example, \texttt{sep(fe simplex) }will first flag any observations
that are perfectly predicted by a single fixed effect and then use
our linear programming-based method to detect additional linear combinations
of regressors that may induce separation. As noted earlier, this approach
will not be able to detect instances where an observation is perfectly
predicted by a combination of multiple fixed effects.

\newcommand{\twoline}[2]{  \begin{tabular}[c]{@{}c@{}}#1\\[2pt]#2\end{tabular}}

\newcommand{\threeline}[3]{  \begin{tabular}[c]{@{}c@{}}#1\\[2pt]#2\\[2pt]#3\end{tabular}}

\begin{table}[htbp]
\caption{Further results from applying different separation checks to BYZ data}
\label{appx-table}
\small
\begin{minipage}{\textwidth}
\begin{center}
\begin{tabular}{llccc}
\hline
\threeline{Stata}{separation}{option} & \threeline{Stata}{absorb}{option} & \threeline{No. of separated}{observations}{detected} & \threeline{Separation}{correctly}{detected?} & \threeline{Estimate}{erroneously}{reported?} \\
\hline
\texttt{sep(none)}&Base&0&No&Yes \tabularnewline
\texttt{sep(ir)}&Base&49&Yes&No \tabularnewline
\texttt{sep(fe)}&Base&42&No&Yes \tabularnewline
\texttt{sep(simplex)}&Base&7&No&No \tabularnewline
\texttt{sep(mu)}&Base&0&No&Yes \tabularnewline
\texttt{sep(fe simplex)}&Base&49&Yes&No \tabularnewline
\texttt{sep(fe simplex)}&Alt.&42&No&Yes \tabularnewline
\texttt{sep(fe ir)}&Base&49&Yes&No \tabularnewline

\hline
\end{tabular}
\end{center}
{\footnotesize These results display the outcomes of what should be equivalent Poisson pseudo-maximum likelihood regressions, applied to the full data set from \citet{baier2019widely}.
In all cases, the model being estimated is given by \eqref{eq:1}.
The Stata command executed for each row is
\texttt{ppmlhdfe flow agr1\_d{*} agr2\_d{*} agr3\_d{*} glob{*}, a(\ldots) sep(\ldots)}.
The option \texttt{sep()} determines the separation checks to be used, and takes the value given in the first column.
The option \texttt{a()} determines the fixed effects to be partialed out; these fixed effects are always \texttt{expcode\#impcode expcode\#year impcode\#year}, indicating exporter-importer, exporter-year, and importer-year fixed effects. However, when the second column is equal to ``Alt.'', the order of the fixed effects is flipped to \texttt{expcode\#year impcode\#year expcode\#impcode} (so the exporter-importer fixed effects are partialed out last). This change should not make any difference but does, thus showcasing the fragility of the simplex method.
}
\end{minipage}
\end{table}

As shown previously in Section \ref{sec:Empirical-Example}, our iterative
rectifier algorithm correctly identifies all 49 separated observations,
while not implementing any checks results in a spurious estimate for
the effect of the EFTA-Romania agreement on Iceland-Romania trade.
Interestingly, though in principle our other methods are applicable
here, they only have mixed success. The \texttt{simplex }method turns
out to be numerically sensitive to the order in which the fixed effects
are encoded in the algorithm.\footnote{This ordering matters because it determines the order in which each
fixed effect is partialed out of each non-fixed effect regressor in
the first step of this procedure that is used to obtain the residuals
to be fed to the linear programming step. If the problematic regressor
is first partialed out with respect to the pair fixed effect, we obtain
a residual that is uniformly zero numerically, as we should. However,
if instead partial out with respect to either of the other fixed effects
first, the computed residual is numerically different enough from
zero that Stata's internal collinearity check algorithm will not flag
it as collinear with the other similarly obtained residuals. } As feared, the \texttt{mu }method fails to detect any of the separated
observations before the algorithm reaches convergence.

\end{document}